\documentclass[a4paper,fleqn,usenatbib]{mnras}

\usepackage{natbib}
\usepackage{soul}

\usepackage{newtxtext,newtxmath}
\usepackage[T1]{fontenc}
\usepackage{ae,aecompl}

\usepackage{todonotes}
\usepackage{outlines}
\usepackage{color}

\usepackage{graphicx}	% Including figure files
\usepackage{amsmath}	% Advanced maths commands
\usepackage{amssymb}	% Extra maths symbols

\usepackage{natbib}

\newcommand{\dd}{\mathrm{d}}
\newcommand{\hmsol}{h^{-1}M_\odot}
\newcommand{\ms}{M_*}
\newcommand{\mh}{M_h}
\newcommand{\mpc}{\mathrm{Mpc}}
\newcommand{\hmpc}{h^{-1}\mathrm{Mpc}}
\newcommand{\cubichmpc}{h^{-3}\mathrm{Mpc}^3}

\newcommand{\cubichgpc}{h^{-3}\mathrm{Gpc}^3}
\newcommand{\hkpc}{h^{-1}\mathrm{kpc}}
\newcommand{\cass}{c_{\mathrm{assem}}}

\newcommand{\hhmsol}{h^{-2}M_\odot}
\newcommand{\ds}{\Delta\Sigma}
\newcommand{\mk}{M_{g{-}r}}
\newcommand{\fred}{f_{\mathrm{red}}(R)}
\newcommand{\kms}{\mathrm{km}\,s^{-1}}

\newcommand{\mhqc}{M_h^{qc}}

\newcommand{\mhqs}{M_h^{qs}}

\newcommand{\muc}{\mu^{c}}
\newcommand{\mus}{\mu^{s}}
\newcommand{\frcen}{f^{\mathrm{red}}_{\mathrm{cen}}}
\newcommand{\frsat}{f^{\mathrm{red}}_{\mathrm{sat}}}
\newcommand{\fbcen}{f^{\mathrm{blue}}_{\mathrm{cen}}}
\newcommand{\fbsat}{f^{\mathrm{blue}}_{\mathrm{sat}}}
\newcommand{\mr}{\mathcal{R}^+}
\newcommand{\lr}{\mathcal{R}^-}
\newcommand{\mb}{\mathcal{B}^+}
\newcommand{\lb}{\mathcal{B}^-}
\newcommand\ihod{\texttt{iHOD}}
\newcommand\bolshoi{\texttt{Bolshoi}}
\newcommand\mdr{\texttt{MDR1}}
\newcommand{\rom}[1]{\uppercase\expandafter{\romannumeral #1\relax}}

\title[Environmental effects induced by halo quenching]{Mapping stellar content to dark matter halos $-$ III.
    Environmental dependence and conformity of galaxy colours}

\author[Zu \& Mandelbaum 2017]{
Ying  Zu$^{1, 2}$\thanks{E-mail: zu.4@osu.edu},
Rachel Mandelbaum$^{2}$
\\
$^{1}$Center for Cosmology and AstroParticle Physics (CCAPP),
Ohio State University, Columbus, OH 43210, USA\\
$^{2}$McWilliams Center for Cosmology, Department of Physics, Carnegie Mellon University, 5000 Forbes Avenue,
Pittsburgh, PA 15213, USA\\
}

\date{Accepted XXX. Received YYY; in original form ZZZ}

\pubyear{2017}

\begin{document}

\label{firstpage}
\pagerange{\pageref{firstpage}--\pageref{lastpage}}
\maketitle

\begin{abstract}
    Recent studies suggest that the quenching properties of galaxies are correlated over several
    mega-parsecs. The large-scale ``galactic conformity'' phenomenon around central galaxies has been
    regarded as a potential signature of ``galaxy assembly bias'' or ``pre-heating'', both of which
    interpret conformity as a result of {\it direct} environmental effects acting on galaxy formation.
    Building on the \ihod\ halo quenching framework developed in~\citet{zu15, zu16}, we discover that
    our fiducial halo mass quenching model, without any galaxy assembly bias, can successfully explain
    the overall environmental dependence and the conformity of galaxy colours in SDSS, as measured by the
    mark correlation functions of galaxy colours and the red galaxy fractions around isolated primaries,
    respectively.  Our fiducial \ihod\ halo quenching mock also correctly predicts the differences
    in the spatial clustering and galaxy-galaxy lensing signals between the more vs.\ less red galaxy
    subsamples, split by the red-sequence ridge-line at fixed stellar mass. Meanwhile, models that tie
    galaxy colours fully or partially to halo assembly bias have difficulties in matching all these
    observables simultaneously. Therefore, we demonstrate that the observed environmental dependence of
    galaxy colours can be naturally explained by the combination of 1) halo quenching and 2) the variation
    of halo mass function with environment --- an {\it indirect} environmental effect mediated by two
    separate physical processes.
\end{abstract}

%%%%%%%%%%%%%%%%%%%%%%%%%%%%%%%%%%%%%%%%%%%%%%%%%%

%%%%%%%%%%%%%%%%% BODY OF PAPER %%%%%%%%%%%%%%%%%%

%%%%%%%%%%%%%%%%%%%%%%%%%%%%%%%%%%%%%%%%%%%%%%%%%%

\vspace{1in}
\section{Introduction}
\label{sec:intro}

Recent studies have shown that the quenching~(i.e., the cessation of star forming activities in galaxies)
properties of central galaxies, such as star formation rate~(SFR), morphology, neutral hydrogen content,
and broad-band colour, are correlated with those of their neighbouring galaxies~\citep{weinmann06, yang06,
ann08, kauffmann10, prescott11, kauffmann13, knobel15, hartley15, wang15, kawinwanichakij16, berti17}.
This so-called ``galactic conformity'' phenomenon exists over two distinct distance scales between the
central galaxy of a primary dark matter halo and its surrounding galaxies, including the true satellite
galaxies within the same halo~\citep[${<}0.5\,\hmpc$; first detected
by][]{weinmann06}\footnote{\citet{weinmann06} compared
    group-sized halos at fixed total optical luminosity, which they used as a proxy for halo mass.},
    and galaxies in halos that are
a few virial radii away from the primary~(${\sim}3\,\hmpc$) --- effects we refer to as ``1-halo'' and
``2-halo'' conformities, respectively. In essence, galactic conformity is a manifestation of some
unknown environmental effect on galaxy formation, and is closely related to the colour-density or
morphology-density relation that has been known for many decades~\citep{oemler74, dressler80}. However,
it is not clear whether galactic conformity extends to even larger scales, and the underlying driver of
this environmental effect remains one of the most important open questions in galaxy formation theory.

The 1-halo conformity is closely related to the physics of galaxy quenching within individual halos. For
instance, virial shocks can heat the incoming gas to high temperatures and inhibit star formation if the halo
is more massive than a few ${\times}10^{12}\hmsol$~\citep{birnboim03, cattaneo06, dekel06, ocvirk08, keres09}.
Massive clusters can even drive extended distribution of hot halo gas via accretion shocks up to several times
the virial radii~\citep{bahe13, gabor15, zinger16}.  Feedbacks from active galactic nucleus~(AGN) can also
potentially truncate the merger-driven star formation episodes at high redshift~\citep{dimatteo2005,
hopkins05}, and keep the hot gas from cooling efficiently at low redshift~\citep{croton06, fabian12,
cheung16}.  In addition, the hot halo gas can strip the gaseous disks from newly accreted
satellites~\citep{gunn72, mccarthy08, font08}, while transforming spirals into S0 galaxies~\citep{bekki02}.
The efficiencies of all those physical processes are directly tied to the halo mass, and we will collectively
refer to them as ``halo quenching''.

Recently, from the weak lensing measurements of the locally brightest galaxies in the Sloan Digital Sky
Survey~\citep[SDSS;][]{york00}, \citet{mandelbaum16} discovered a strong bimodality in the average host halo
mass of the red vs.\ blue central galaxies --- at fixed $\ms$, red central galaxies preferentially live in
halos that are factor of two~($\ms\,{\sim}\,2{\times}10^{10}\,\hhmsol$) to almost
ten~($\ms\,{>}\,2{\times}10^{11}\,\hhmsol$)
more massive than ones that host blue centrals~\citep[see also][]{mandelbaum06, more11}. \citet{zu16}
interpreted this halo mass bimodality as a pronounced signature of halo quenching, and demonstrated that
models without an explicit halo quenching are unlikely able to reproduce such strong bimodality in halo mass.
For example, in a model that maximizes the so-called ``galaxy assembly bias''~\citep{zhu06, croton07, zu08, zentner14} by matching galaxy colours to
the formation time of halos~\citep[i.e., age-matching;][]{hearin13}, blue central galaxies are instead placed
in slightly more massive halos than red centrals, due to the weak anti-correlation between the formation time
and mass of halos.

The observed 1-halo conformity effects are consistent with the halo quenching scenario.  \citet{wang12}
found qualitative agreement between the 1-halo colour conformities observed in SDSS and predicted by
the~\citet{guo11} semi-analytic model~(SAM), where star formation is regulated by the AGN ``radio-mode''
feedback with an efficiency that is directly tied with halo mass $\mh$~\citep{croton06}.  Using morphology
as the quenching indicator, \citet{ann08} found a similar 1-halo conformity effect in SDSS and argued
that the hot halo gas in high-mass systems could be responsible for the coherent transformation of
galaxy morphologies.

In the 2-halo regime, a possible conformity effect was first detected by \citet{kauffmann10} using
photometrically-selected satellite galaxies in SDSS. Switching to the spectroscopic satellite sample,
\citet{kauffmann13} found that for primaries with $\ms$ around a few ${\times}\,10^{10}\,\hhmsol$, strong
conformity between the gas-poor primaries and the HI content of satellites persists at projected distances of
${\sim}\,3\,\hmpc$, where a direct causal link between the two becomes unlikely; for primaries with
$\ms\,{\sim}\,10^{11}\,\hhmsol$, the large-scale conformity signal is weak and confined within a couple virial radii
of clusters.

One potential problem with the 2-halo conformity detection is, the primary galaxies in \citet{kauffmann13}
were selected by an isolation criteria that could include gas-poor low-mass central galaxies within the
vicinity of a massive companion~\citep{stark16}, or even mis-identify a small fraction of satellites within
a larger system as primaries~\citep{tinker17}.  \citet{kauffmann13} explored this contamination issue
using the \citet{guo11} SAM mock catalogue, and argued that the impact is too small to account for the
observed conformity. However, using a group finding algorithm for identifying centrals, \citet{tinker17}
argued that the 2-halo conformity seen by \citet{kauffmann13} could be artificially boosted by the
mis-identified primaries in the sample.  \citet{sin17} also discovered that the isolation criteria of
\citeauthor{kauffmann13} could include low-mass central galaxies in the vicinity of massive systems,
and that the large-scale conformity signal is likely a short-range effect sourced by massive halos.

Currently, there are two possible explanations of the 2-halo conformity effect. \citet{kauffmann13} argued
that the strong signal around the very low-mass central galaxies~(a few ${\times}10^{9}\,\hhmsol$) favors the
``pre-heating'' scenario in which large reservoirs of intergalactic medium~(IGM) was heated up to a high
entropy level, probably due to bursty star-forming activities at early epochs or spatially-coherent injection
of energy from AGN/stellar feedbacks~\citep{mo02}. Despite the lack of robust detections so
far~\citep[see][]{yang06, lin16, vakili16}, the galaxy assembly bias effect can also produce a strong 2-halo
conformity signal, by making use of the coupling between halo accretion histories in the same density
environment across large scales~\citep{hearin15}.  In essence, both pre-heating and galaxy assembly bias are
{\it direct} environmental effects on two key parameters of galaxy formation, i.e., the entropy of IGM and the
overall accretion rate of baryons, respectively.

Alternatively, an {\it indirect} environmental effect, such as the combination of the environmental
dependence of halo mass function and the simple halo quenching mechanism, should also give rise to a
large-scale galactic conformity. This third, indirect effect has not been adequately explored in previous
studies, partly due to a common mis-conception that there is no inherent environmental dependence in
the standard halo model framework~\citep{cooray02}. But as articulated by~\citet{martino09}~\citep[see
also][]{mo96, lemson99, sheth02}, the dependence of halo abundance and formation history on the
large-scale density environment is a standard element of the excursion set theory of cosmological
structure formation~\citep{press74, bond91, lacey93}, which also allows mass-dependent biasing to be
understood within the peak-background split formalism~\citep{bardeen86, sheth99}.

Is halo quenching consistent with the environmental dependence of colours observed in SDSS?
In the pioneering work by~\citet{skibba09}, they demonstrated that a simple Halo Occupation
Distribution~\citep[HOD;][]{jing98, ma00, peacock00, seljak00, yang03, scoccimarro01, berlind02, guzik02,
zheng05, mandelbaum06, vandenbosch07, ross09, leauthaud12, zu15} model of galaxy colours, without assembly bias,
is able to reproduce the level of environmental dependence of colours in SDSS up to $20\,\hmpc$. They are
the first to employ the mark correlation functions of colours as a robust measure of the environmental
dependence of galaxy quenching, which includes contributions from the large-scale conformity around
central galaxies and the correlation between the colours of satellites inside different halos. However,
\citet{skibba09} made a few overly-simplified assumptions, including that the galaxy colour depends solely
on luminosity. As a result, this model does not represent a viable halo quenching model, and is thus
unlikely able to explain the observed strong halo mass bimodality of central galaxies in SDSS~\citep{zu16}.

Similarly, \citet{hearin15} constructed an HOD mock of galaxy colours that has the similar $\mh$-dependence of
red galaxy fractions as their age-matching mock, but without any galaxy assembly bias.  They found that this
HOD mock does not produce any conformity effects, as measured by the red galaxy fractions around the red vs.
blue primaries selected in the same way as in \citet{kauffmann13}.  \citet{hearin15} further argued that a
large-scale conformity signal is the smoking-gun evidence of galaxy assembly bias.  However, despite the
technical differences between the HOD mocks built by \citet{skibba09} and \citet{hearin15}, it is quite
intriguing that a strong overall environmental dependence of colours seen in one may not yield an equally strong
conformity signals of centrals in the other.

Therefore, it is very important to explore whether a viable halo quenching model within the HOD framework
can simultaneously explain the environmental effects in the spatial distribution of galaxy colours and the
large-scale colour conformity signals observed in SDSS.  In this paper, we build on the best-fitting halo
quenching model within the \ihod\ framework developed in our Paper I~\citep{zu15} and II~\citep{zu16}
of this series, and employ the mark correlation functions and the red galaxy fractions at fixed $\ms$
as the joint probe of environmental effects in our analysis. To better distinguish different models of
galaxy colours, we also investigate the spatial clustering and the g-g lensing of more vs.\ less red
galaxies, split by the red-sequence~(RS) ridge-line at fixed $\ms$.

This paper is organized as follows. We briefly describe the \ihod\ framework and the simple halo quenching
model in \S~\ref{sec:oldihod}, and introduce our three colour assignment schemes in \S~\ref{sec:quenching}.
We present our main findings in \S~\ref{sec:results} and conclude by summarizing our results and looking to
the future in \S~\ref{sec:sum}.

\section{The {\large \ihod} Model and Mock Galaxy Catalogues}
\label{sec:oldihod}

The mock galaxy catalogues in this study are built on the \ihod\ model developed in Papers I \& II.  We will
briefly describe the main features of \ihod\ below, and refer readers to Papers I \& II for more details,
including the mathematical framework of the model, the selection of galaxy samples from SDSS, and the
measurement of $w_p$ and $\ds$ for those samples. We will also describe the $N$-body simulations and
procedures we employ to generate the mock galaxy catalogues. Readers who are familiar with Papers I \& II can
skip the next subsection and start from~\S~\ref{subsec:mocks}.

For the conformity ``mark'' we focus on the $g{-}r$ colours~($K$-corrected to $z{=}0.1$), which are measured
much more robustly than other quenching indicators like the SFR or HI gas mass. Studies of environmental
effects are particularly sensitive to the volume size, and \citet{xu16} demonstrated that the clustering of
low-$\ms$ galaxies in the local volume below $z{=}0.03$ are subjected to very severe cosmic variance effect.
Therefore, we limit our analyses to galaxies with $\ms{\geqslant}10^{10}\,\hmsol$, so that the minimum
redshift range of our volume-limited galaxy samples is $z{=}[0.01, 0.07]$, i.e., the redshift range of our
lowest-$\ms$ sample. The maximum redshift probed by the high-$\ms$ samples is $0.2$, and the
median redshift of the SDSS galaxies used this analysis is around $0.1$.

Throughout this paper and Papers I \& II, all the length and mass units are scaled as if the Hubble constant
were $100\,\kms\mpc^{-1}$. In particular, all the separations are expressed in co-moving distances in units of either
$\hkpc$ or $\hmpc$, and the stellar masses and halo masses are in units of $\hhmsol$ and $\hmsol$, respectively.
We employ the stellar mass estimates from the latest MPA/JHU value-added galaxy
catalogue\footnote{\url{http://home.strw.leidenuniv.nl/~jarle/SDSS/}}, and convert other stellar
masses~\citep[e.g., in the age-maching mock of ][]{hearin14} to the MPA/JHU values using the fitting formulae given by~\citet{li09}.
Unless otherwise noted, the halo mass is defined by
$\mh\,{\equiv}\,M_{200m}\,{=}\,200\bar{\rho}_m(4\pi/3)r_{200m}^3$, where $r_{200m}$ is the corresponding halo
radius within which the average density of the enclosed mass is $200$ times the mean matter density of the
Universe, $\bar{\rho}_m$.

\subsection{The {\large \ihod} Model and Simple Halo Mass Quenching}
\label{subsec:oldihod}

The \ihod\ framework aims to describe the {\it probabilistic} connection between galaxies and halos, assuming
that the enormous diversity in the individual galaxy assembly histories inside similar halos would reduce to a
stochastic scatter about the {\it mean} galaxy-to-halo relation by virtue of the central limit theorem.
Therefore, the key is to derive $P(\mathbf{g} \mid \mathbf{h})$, the conditional probability distribution
function~(PDF) of galaxy properties $\mathbf{g}$ at fixed halo properties $\mathbf{h}$, where $\mathbf{g}$ and
$\mathbf{h}$ are the corresponding vectors that describe the most important sets of galaxy and halo
properties. For example, we could include stellar mass, optical colour, SFR, and morphology in $\mathbf{g}$,
and halo mass, concentration, spin, and tidal environment in $\mathbf{h}$.

In Paper I and II, we have applied the \ihod\ model to the SDSS main spectroscopic sample, and successfully
mapped the red and blue galaxies at different stellar masses to their underlying halos. In this first-cut
analysis, we adopted a binary colour variable $b_{g{-}r}$, by classifying each galaxy into either
red~($b_{g{-}r}{=}1$) or blue~($b_{g{-}r}{=}0$) based on a $\ms$-dependent colour-split
\begin{equation}
\left(g-r\right)_{\mathrm{split}}|_{\ms} = 0.8 \left(\frac{\lg\ms}{10.5}\right)^{0.6}.
\label{eqn:cut}
\end{equation}
We then derived the conditional probability distribution of halo mass for galaxies at fixed stellar mass and
colour category $P(\mh\,|\,M_*,\, b_{g{-}r})$, from the stellar mass and color dependence of galaxy
clustering~($w_p$) and g-g lensing~($\ds$) measurements. Thanks to the probabilistic nature of the model, we
are able to include ${\sim}80\%$ more galaxies in the analysis than traditional HOD methods, while accounting for
the incompleteness of galaxy samples in a statistically consistent fashion.

In practice, we first derive the overall stellar-to-halo connection $P(\mh\,|\,M_*)$ in Paper I from the
stellar mass dependence of $w_p$ and $\ds$. In Paper II, we describe the halo quenching effect statistically
using the red galaxy fractions of centrals and satellites as functions of $\mh$~(c.f. equations 12 and 13 of
Paper II),
\begin{equation}
    \frcen(\mh) = 1 - \fbcen(\mh) = 1 - \exp\left[-\left( \mh/\mhqc\right)^{\muc}\right],
    \label{eqn:frcenhalo}
\end{equation}
and
\begin{equation}
    \frsat(\mh) = 1 - \fbsat(\mh) = 1 - \exp\left[-\left( \mh/\mhqs\right)^{\mus}\right],
    \label{eqn:frsathalo}
\end{equation}
where $\mhqc$ and $\mhqs$ are the critical halo masses responsible for triggering quenching of central and
satellites, respectively, and $\muc$ and $\mus$ are the respective powered-exponential indices controlling the
transitional behavior of halo quenching across the critical halo masses. Therefore, by combining
Equations~\ref{eqn:frcenhalo} and~\ref{eqn:frsathalo} together with the overall $P(\mh \mid M_*)$, we now
arrive at a complete model for $P(\mh\,|\,M_*,\,b_{g{-}r})$.

More importantly, the best-fitting $P(\mh\,|\,M_*,\,b_{g{-}r})$ successfully predicts the strong bimodality in
the host halo mass distributions of the red and blue galaxies in SDSS~\citep{mandelbaum16}, which implies a
dominant halo quenching mechanism that turns on in halos above
$\mhqc{\simeq}\mhqs{\simeq}1.5\times10^{12}\,\hhmsol$~(with different powered-exponential indices for central
and satellites).  This success is highly non-trivial, as many alternative models that strive to explain galaxy
colours fail this test~\citep[e.g., by assigning galaxy colours based on halo age or stellar mass;][]{zu16}.

One of the interesting extensions of the current \ihod\ model, expressed by $P(\mh\,|\,M_*,\,b_{g{-}r})$, is
to add an important secondary halo property, such as concentration $c$, to see whether it would provide a more
comprehensive description of the observed galaxy colours, especially when $g{-}r$ is included as a continuous
variable instead of a binary one. This extension, expressed by $P(\mh,\,c \,|\,\ms,\, g{-}r)$, also represents
a useful formalism for including galaxy assembly bias when connecting galaxy colours to halos, because
concentration is one of the best indicators for halo assembly bias~\citep[see also][]{paranjape15, puebla16,
    hearin16b, pahwa16, puebla17}.

\subsection{From {\large \ihod} to Mock Galaxy Catalogues}
\label{subsec:mocks}

\begin{figure*}
\begin{center}
    \includegraphics[width=1.0\textwidth]{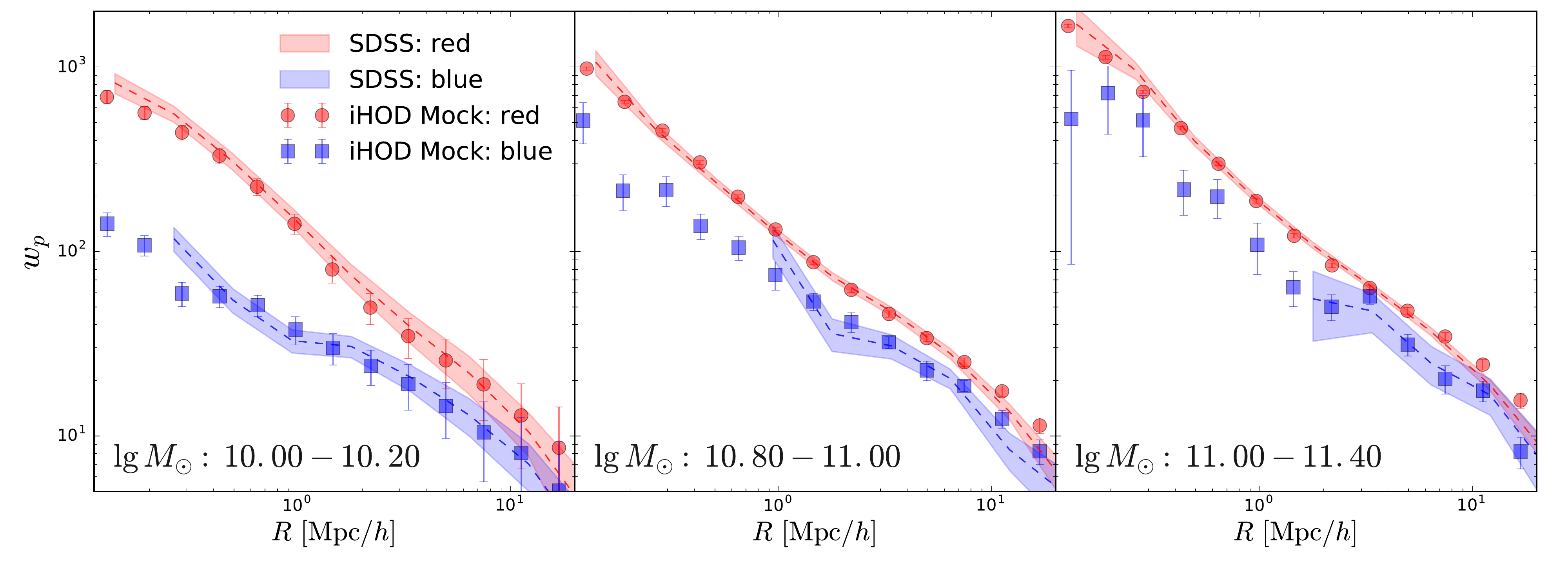} \caption{Comparison between
    the projected auto-correlation functions $w_p(R)$ measured from the SDSS main sample~(dashed curves with
    shaded uncertainty bands) and predicted by the \ihod\ mock galaxy catalogues with halo quenching~(data
    points with errorbars), for red and blue galaxies within three different stellar mass bins~($\lg\ms$
    range listed in the bottom left of each panel), respectively. The low~(left panel) stellar mass
    mock sample is derived from \bolshoi, while the high~(right panel) and intermediate~(middle panel)
    stellar mass samples are from \mdr. The blue bands are cut off on small scales because the SDSS blue
    galaxies becomes increasingly rare at higher $\ms$.}
\label{fig:wpcolor}
\end{center}
\end{figure*}

In Papers I \& II we adopted an analytic method to predict the projected galaxy correlation function $w_p$
and g-g lensing signal $\ds$ at fixed $\ms$ for the red and blue galaxies in SDSS. In this paper, a fully
analytic approach will not enable us to easily address the question of interest,
so we will predict $w_p$, $\ds$, as well as the colour-mark
correlation function $\mk$ and the red galaxy fraction as a function of projected distance $\fred$ using
direct measurements from the mock galaxy catalogues generated under \ihod~\citep[but see][for an analytic
halo-model description of $\mk$]{sheth05b}.

In order to cover a dynamic range of two orders of magnitude in stellar mass~($10^{10}$-$10^{12}\,\hhmsol$),
we employ two cosmological {\it N}-body simulations that are evolved from the same {\it WMAP5} cosmological
parameters~\citep{komatsu09}, but with complementary sets of mass resolutions and volumes. For mock galaxies
with $\lg\ms\,{<}\,10.6$, we use the \texttt{Bolshoi}~\citep{klypin11} simulation because of its higher mass
resolution~($1.35\,{\times}\,10^8\,\hmsol$). The volume of \texttt{Bolshoi} is too small~($250^3\,\cubichmpc$)
to overcome the cosmic variance of more massive galaxies, so we use the \texttt{MDR1}
simulation~\citep[$8.721\,{\times}\,10^9\,\hmsol$ particle mass and a $1\,\cubichgpc$ box;][]{prada12} for
deriving mock galaxies with $\lg\ms\,{\geq}\,10.6$. In both simulations, we make use of the halo catalogues
identified by the \texttt{ROCKSTAR}~\citep{behroozi13} spherical overdensity halo finder at $z\,{=}\,0.1$, the
median redshift of our SDSS galaxy samples.

Since the cosmology assumed in Papers I \& II is slightly different from the {\it WMAP5} values used by the
simulations, we have re-calculated the constraints on the global \ihod\ parameters and the halo quenching
model using the {\it WMAP5} cosmology. We have also updated our analytic model in Paper I by including the
so-called ``residual redshift-space distortion'' effect in $w_p$, using the correction method described in
\citet{vandenbosch13}. The main change in cosmology is the increase of $\sigma_8$, which we anticipate to
affect primarily $\delta$, the high-mass end slope of the mean stellar-to-halo mass relation~(SHMR),
but cause very little change to the slope at the low-mass end or the scatter about the
mean SHMR~\citep{leauthaud12, zu15}.
Therefore, we only vary $\delta$ during the new fit, while keeping other \ihod\ parameters unchanged from the
constraints listed in the table 2 of Paper I. The new best-fitting $\delta$ is $0.44$~($0.42$ in Paper I).  The
new best-fitting halo quenching parameters are
$\{\lg\mhqc,\,\muc,\,\lg\mhqs,\,\mus\}\,=\,\{11.78,\,0.41,\,12.19,\,0.24\}$, slightly different from Paper II~(c.f., table 2).

Conceptually, \ihod\ mock galaxies can be generated from simulated halo catalogues by drawing $\ms$ and
$b_{g{-}r}$ jointly from $P(M_*,\, b_{g{-}r} \, | \, \mh)$, which can be trivially derived from $P(\mh \, | \,
M_*,\, b_{g{-}r})$ using Bayes' Theorem. But since the centrals and satellites follow distinct stellar-to-halo
mass relations, in practice it is more convenient to assign stellar masses to the centrals and satellites
separately, then label them red or blue, and at last give them positions and velocities. We describe the three
steps in turn below.

\begin{itemize}
    \item The first step is to assign stellar masses. Using the best-fitting \ihod\ parameters as input, we
    derive the mean SHMR and its logarithmic scatter for central
    galaxies~(c.f., Fig. 10 of Paper I), and the conditional stellar mass functions~(CSMF) for satellites
    in halos of different masses~($\phi(\ms^{\mathrm{sat}}\,|\,\mh)$; c.f., Fig. 12 of Paper I).
    For each simulated halo with mass $\mh$, we randomly draw the stellar mass of its central galaxy from the
    log-normal distribution $P(\ms^{\mathrm{cen}}\,|\,\mh)$ specified by the combination of SHMR and its
    logarithmic scatter at $\mh$; we then assign it a set of satellite galaxies, whose stellar
    mass distribution follows $\phi(\ms^{\mathrm{sat}}\,|\,\mh)$.

    \item Secondly, for each central~(satellite) galaxy residing in a halo of $\mh$, we label it red or blue
    according to the mean red central~(satellite) galaxy fraction at that halo mass~(i.e.,
    Equations~\ref{eqn:frcenhalo} and~\ref{eqn:frsathalo}), predicted by the best-fitting halo quenching
    parameters. The label indicates whether the galaxy colour is above or below
    $\left(g-r\right)_{\mathrm{split}}$, but without a particular $g{-}r$ value.

    \item We predict the relative positions and velocities of mock galaxies with respect to the halo center
    as follows. For each main halo with concentration $c$, radius $r_{200m}$, and 3D dark matter velocity
    dispersion $\sigma$, the central galaxy is placed at the center, while the satellite positions are
    assigned randomly according to an isotropic NFW profile with $c_g\,{\equiv}\,f_c \,\times\, c$ and
    a cut-off at $r_{200m}$. The galaxy velocities are assigned based on the galaxy velocity bias model
    described in \citet{guo15}, where the relative velocities of central and satellite galaxies follow
    Gaussian distributions with zero means and standard deviations of $\sigma_c\,{=}\,\alpha_c\,\sigma$ and
    $\sigma_s\,{=}\,\alpha_s\,\sigma$, with $\alpha_c\,{=}\,0.20$ and $\alpha_s\,{=}\,1.00$, respectively.
    Note that we do not fit the observed monopole and quadrupole of the correlation functions for the
    values of $\alpha_s$ and $\alpha_c$, as was done in \citet{guo15}, because our analyses focus on
    projected quantities and the impact of peculiar velocities on those quantities is minimal.
\end{itemize}

It is worth nothing that by adopting a halo boundary of $r_{200m}$, we have implicitly assumed that the halo
quenching effects have a sharp transition across $r_{200m}$ without extending beyond individual halos in those
mocks, which is not necessarily an unreasonable assumption~\citep{baxter17}.  We have also ignored the radial
segregation of satellite stellar mass~\citep{adami98,
    presotto12, roberts15, contini15, kafle16, vandenbosch16, nascimento17} and colour~\citep{chen08,
    prescott11, woo13, wang14, woo17} in this study. Whether a true 2-halo environmental dependence or galactic
conformity would emerge on scales beyond $3\,\hmpc$ is independent of the choice for the halo quenching
boundary or the segregation effects, as they would only modify the shape of the mark correlation functions and
conformity signals in the 1-halo regime.

% We will nonetheless come back to this limitation of our mock catalogues in \S\ref{sec:results}.

As a sanity check, Figure~\ref{fig:wpcolor} compares the projected correlation functions $w_p$ measured from
the SDSS~(dashed curves with shaded uncertainty bands) and the \ihod\ mock galaxy samples~(data points with
errorbars), for red and blue galaxies within three different stellar mass bins.  Clearly, the mock galaxy
samples provide an excellent description of the spatial clustering of the red and blue galaxies in SDSS. In
addition, the g-g lensing signals~(not shown here, but see Fig. 8 in Paper II) predicted by the red and
blue mock galaxies also agree with that measured from SDSS very well.
Taking advantage of the realistic red
galaxies in the mock catalogue, \citet{zu16b} constructed a mock cluster catalogue that mimics the clustering
of the redMaPPer clusters~\citep{rykoff14} and predicted the observed level of cluster assembly bias in SDSS.

We emphasize that it is imperative for the red and blue mock galaxies to accurately reproduce the observed
auto-correlation~($w_p$) and cross--correlation with the dark matter~($\ds$), as demonstrated by
Figure~\ref{fig:wpcolor}. For any mock galaxy catalogues that fail to recover $w_p$ and $\ds$, systematics
discrepancies in these two observables could potentially propagate into the systematic uncertainties of the
mark correlation functions and the red galaxy fractions around isolated primaries, rendering the physical
interpretation of those measurements inconclusive.

\section{Assigning Galaxy Colours}
\label{sec:quenching}

The simple halo quenching model described above allows us to label each mock galaxy as red or blue, i.e., to
predict whether the $g{-}r$ colour of that galaxy is above or below $\left(g-r\right)_{\mathrm{split}}$.
Although this binary split is adequate for the purpose of distinguishing different quenching models in Paper
II, it is insufficient for studying the environmental dependence of colours, which are sensitive to the small yet spatially
coherent variations of galaxy colours. Therefore, we need to further assign specific $g{-}r$ values to the
mock galaxies before comparing them to the data.

Most importantly, we need to make sure that the colour distribution of mock galaxies at any given stellar
mass, $P(g{-}r\,|\,\ms)$, is consistent with SDSS. \citet{kauffmann15} found a significant discrepancy between
the conformity observed in SDSS and that predicted by a cosmological hydrodynamic
simulation~\citep[\texttt{Illustris};][]{vogelsberger14, bray16}; But since the detailed SFR and colour
distributions within \texttt{Illustris} are quite different from those in the observations, it is unclear whether the
existing physical processes in the simulation should be capable of reproducing the correct large-scale
conformity, or some key environmental quenching mechanism is missing. In this analysis, we aim to eliminate this ambiguity
by drawing conclusions based on mock galaxies that are generated with different quenching physics but have the
identical $P(g{-}r\,|\,\ms)$ as the real galaxies.

In this section, we will derive $P(g{-}r\,|\,\ms)$ by fitting double-Gaussian PDFs to SDSS colour
distributions in~\S~\ref{subsec:bimodal}, and comment on the origin of scatter in galaxy colours in
~\S~\ref{subsec:scatter}. We next describe the three colour assignment schemes in turn in
\S~\ref{subsec:scheme} and examine the theoretical sources of conformity in those schemes in
~\S~\ref{subsec:source}.

\subsection{Modelling the Bimodal Distributions of Colour}
\label{subsec:bimodal}

\begin{figure*}
\begin{center}
    \includegraphics[width=1.\textwidth]{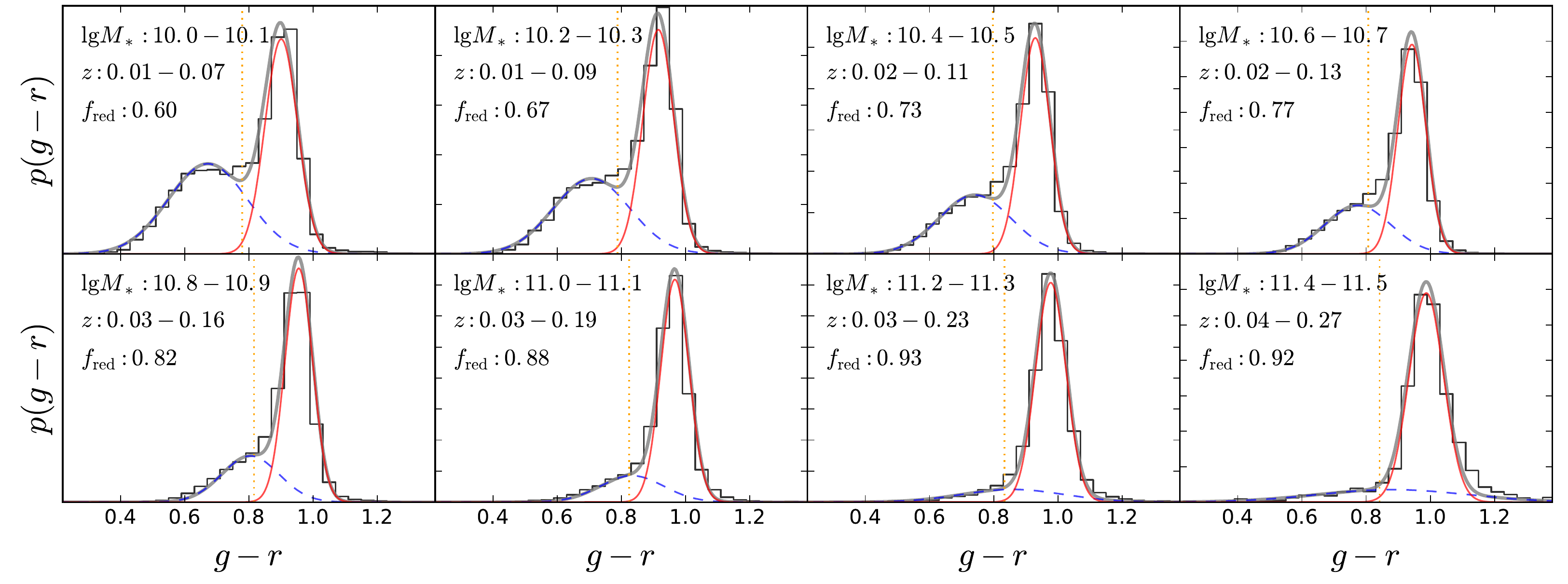}
    \caption[]{Illustration of the $g{-}r$ colour distributions in eight
    different stellar mass bins. In each panel, the black histogram is directly
    measured from a volume-limited SDSS galaxy sample~($\lg\ms$ and $z$ ranges
    of each sample listed in the top left). The gray distribution is the
    best-fitting mixture model that consists of two Gaussian components~(red
    solid and blue dashed). The orange dotted vertical line indicates the
    $\ms$-dependent colour cut used for dividing the galaxies into red and blue
    populations, with the fraction of red galaxies $f_{\mathrm{red}}$ listed in
    the legend.} \label{fig:bimodality}
\end{center}
\end{figure*}

To accurately measure the underlying colour distributions in SDSS, we select a suite of volume-limited stellar
mass-binned samples, with equal bin width of $0.1\,\mathrm{dex}$ starting at $\lg\ms=10.0$. The stellar mass
limit adopted in this study is the same as the ``mixture limit'' $M_*^{\mathrm{mix}}(z)$ defined in Paper I,
\begin{equation}
    \lg M_*^{\mathrm{mix}}(z) = 5.4 \times (z-0.025)^{0.33} + 8.0,
    \label{eqn:ql}
\end{equation}
corresponding to the characteristic stellar mass at which the average galaxy colour $\langle g{-}r \rangle$
sharply transitions from below to above $0.8$ at any given $z$. As explained in Paper I, galaxy samples
selected above this limit can be regarded as being approximately volume-complete. As a result, we measure the
colour distributions from the suite of stellar-mass binned samples, and use them directly as the input data
for fitting $P(g{-}r\,|\,\ms)$.

Following previous studies of colour bimodality~\citep{skibba09, taylor15, paranjape15}, we employ the
double-Gaussian function as the analytic form of $P(g{-}r\,|\,\ms)$,
\begin{equation}
    P(g{-}r\mid\ms) = f^{\prime}_{\mathrm{red}}(\ms) \,
    \mathcal{N}^{\mathrm{red}} + [1-f^{\prime}_{\mathrm{red}}(\ms)] \, \mathcal{N}^{\mathrm{blue}},
\end{equation}
where $\mathcal{N}^{\mathrm{red}}$ and $\mathcal{N}^{\mathrm{blue}}$ represent the red and blue Gaussian
components, respectively, while $f^{\prime}_{\mathrm{red}}(\ms)$ is the fraction of the red Gaussian
component.  Note that $f^{\prime}_{\mathrm{red}}(\ms)$ is an unknown free parameter in the double-Gaussian model,
different from the usual red galaxy fraction
\begin{equation}
    f_{\mathrm{red}}(\ms) = \int_{(g-r)_{\mathrm{split}}|_{\ms}}^{+\infty} P(g{-}r\mid\ms)\,\dd\,(g{-}r),
\end{equation}
which can be directly measured from SDSS.
During the fitting, we not only minimize the $\chi^2$ between the double-Gaussian and the colour distribution
measured from each volume-limited sample, but also ensure that the predicted $f_{\mathrm{red}}$ is equal to
the measured value from SDSS.
We impose the latter primarily for the sake of consistency between the colour
assignments below and the halo quenching model described in \S~\ref{sec:oldihod}, which is based on the
division of red vs.\ blue galaxies using $(g-r)_{\mathrm{split}}$.

Figure~\ref{fig:bimodality} compares the best-fitting double-Gaussians~(gray curves) with the colour
distributions~(black histograms) measured from eight selected stellar mass samples, with the stellar mass
range, redshift range, and the red galaxy fraction of each sample indicated on the top left of every panel.
The red solid and blue dashed distributions indicate the red and blue Gaussian components, and the orange
dotted vertical lines indicate the red vs.\ blue division defined by $(g-r)_{\mathrm{split}}$. In general, the
best-fitting models of $P(g{-}r\,|\,\ms)$ provide an adequate description of the observed colour distributions
across all measured stellar mass bins, showing only very minor discrepancies in the so-called ``green valley''
where the two components overlap.

\begin{figure}
\begin{center}
    \includegraphics[width=0.4\textwidth]{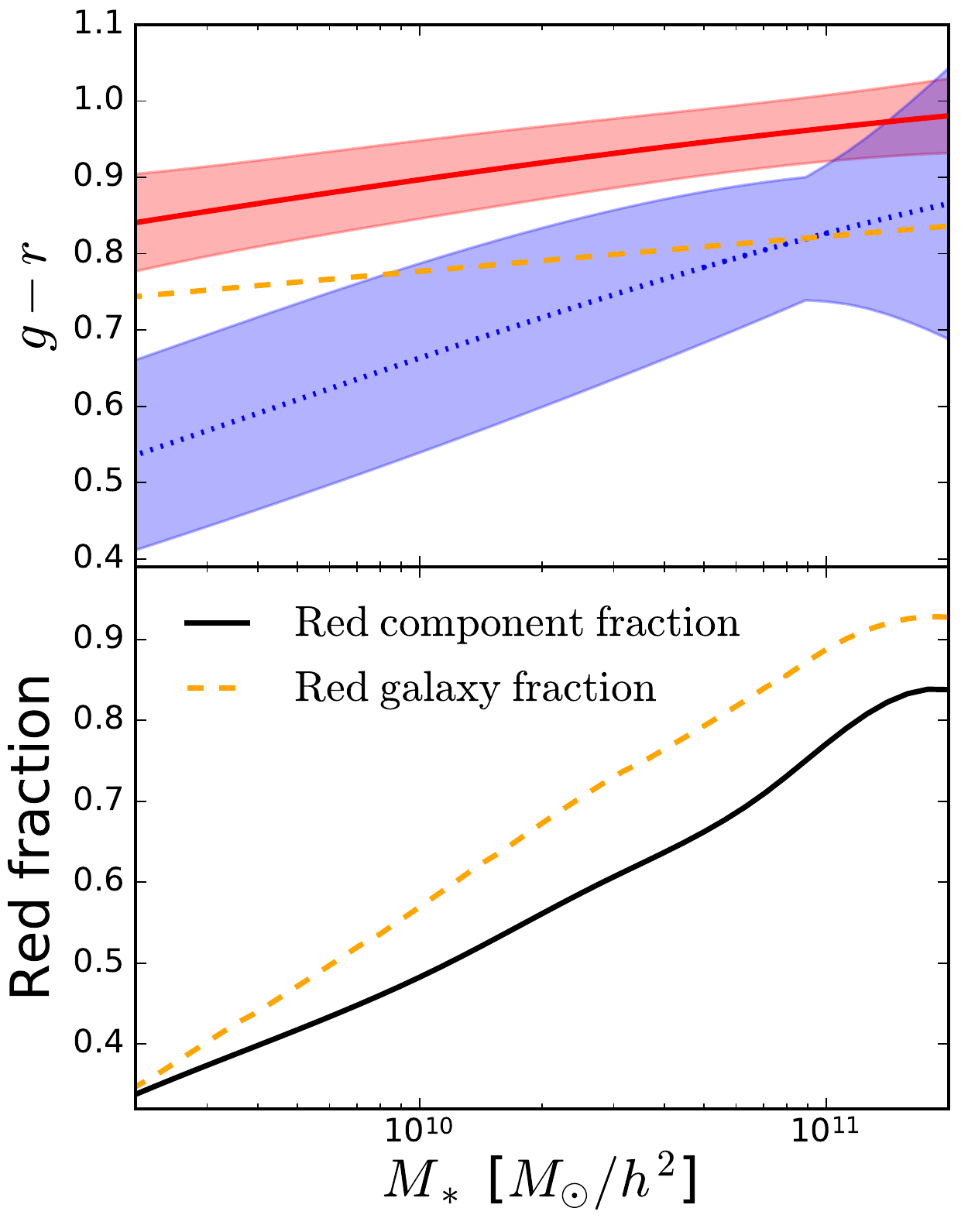} \caption[]{{\it Top panel}: The derived
	colour-stellar mass relations of the red and blue Gaussian components. The red solid curve and the
	width of the red shaded band indicate the mean and the dispersion of the red component,
    while the blue dotted curve and shaded band are the blue component. The orange dashed line is the
    colour division we use to separate galaxies into red and blue populations~(Equation~\ref{eqn:cut}).
	{\it Bottom panel}: The black solid curve is the fraction of the red Gaussian component as a function
	of $\ms$, using the normalization of the red Gaussian component at each $\ms$; the orange dashed curve
	is the fraction of the red galaxies as a function of $\ms$, using the orange dashed line in the top
    panel to split galaxies into red and blue.} \label{fig:fred}
\end{center}
\end{figure}

Figure~\ref{fig:fred} summarizes the result of the fitting. In the top panel, the red solid and blue dotted
lines indicate the means of the red and blue Gaussian components, respectively, while each colour-shaded band
represents the scatter about the respective mean. The orange dashed line indicates $(g-r)_{\mathrm{split}}$,
the red vs.\ blue division used for defining $f_{\mathrm{red}}$. Clearly, there is progressively more overlap
between the two Gaussian components at higher mass, mainly due to the two means approaching each other. As a
result, there is a discrepancy of ${\sim}\,0.10$ between $f_{\mathrm{red}}$ and $f^{\prime}_{\mathrm{red}}$
above $\lg\ms\,{=}\,10$, as shown by the two curves in the bottom panel. The strong overlap makes it
difficult to distinguish the true quiescent vs.\ active galaxies based on colour at the high-mass
end~(discussed further later).

We will assign galaxy colours in two steps. In the first step, we generate an ensemble of mock colours at each
$\ms$~(adopting a bin size of $\Delta\lg\ms\,{=}\,0.1$) by drawing random $g{-}r$ values from the best-fitting
analytic $P(g{-}r\,|\,\ms)$.  Within each narrow stellar mass bin, we divide those mock colours into red and
blue, so that the colours of red galaxies follow
\begin{eqnarray}
P^{\mathrm{red}}(g{-}r\mid\ms)&=&\left\{
\begin{array}{ll}
  \frac{P(g{-}r\mid\ms)}{f_{\mathrm{red}}(\ms)} & \quad\mbox{if $g{-}r{\geqslant}(g{-}r)_{\mathrm{split}}$ }\\
 0 & \quad\mbox{if $g{-}r{<}(g{-}r)_{\mathrm{split}}$ } ,
\end{array} \right.
\label{eqn:pred}
\end{eqnarray}
while the colours of blue galaxies follow
\begin{eqnarray}
P^{\mathrm{blue}}(g{-}r\mid\ms)&=&\left\{
\begin{array}{ll}
  \frac{P(g{-}r\mid\ms)}{1-f_{\mathrm{red}}(\ms)} & \quad\mbox{if $g{-}r{<}(g{-}r)_{\mathrm{split}}$ }\\
 0 & \quad\mbox{if $g{-}r{\geqslant}(g{-}r)_{\mathrm{split}}$ } .
\end{array} \right.
\label{eqn:pblue}
\end{eqnarray}
We then distribute those red~(blue) colours among the red~(blue) galaxies based on three different schemes
below.  In this way, a red~(blue) galaxy will retain its red~(blue) label across the three catalogues, but
obtain a different red~(blue) $g{-}r$ colour depending on the relative strength of halo quenching vs.\ galaxy
assembly bias assumed in each catalogue.

\subsection{Origin of Scatter in the Colour-Stellar Mass Relations}
\label{subsec:scatter}

In essence, our goal is to find out which halo properties are the most responsible for driving the intrinsic
scatter in the colour--stellar mass relation~(CSMR) of red or blue galaxies. Below $10^{11}\hhmsol$, the
scatter in the red/blue galaxy CSMR at any given $\ms$ is dominated by that of the intrinsic stellar colours
of galaxies, which reflects the diversity in the integrated star-formation histories of those red/blue
galaxies at that stellar mass. Starting from $\lg\ms{=}11$, however, there is a sudden increase in the scatter
of the blue component, as shown by the blue shaded band in the top panel of Figure~\ref{fig:fred}. This
increase is unlikely intrinsic, but caused by a combination of strong dust reddening in some edge-on spirals
and starbursts triggered by major mergers. The internal dust reddening in spirals only becomes prominent
at the high mass end, probably because they have accumulated significant dust during past star formation
activities and are physically large with long path lengths through the edge-on discs~\citep{masters10}.

Ideally, we would prefer using the intrinsic stellar colours that are corrected for internal dust.
\citet{taylor15} tried to estimate the internal dust reddening empirically from the internal
extinctions~($A_V$) given by Stellar Population Synthesis fits~\citep[assuming the extinction law
of][]{calzetti00}. After subtracting this estimated reddening, they found substantial decrease in the
dispersion of the blue component at high $\ms$, mostly due to the large corrections to dusty star-forming systems.
However, it is quite difficult to assess the systematic uncertainties associated with the assumed dust
properties, especially the propensity to over-correct systems with little or no dust~\citep{taylor15}.
Therefore, for the sake of simplicity, we will not apply any reddening correction on the $g{-}r$ colours in
this analysis, and assume that the extrinsic scatter can be effectively accounted for by reducing the
cross-correlation
between the observed colour and halo properties. For galaxies with $\ms\,{>}\,10^{11}\,\hhmsol$, the
cross-correlation is further diluted by the extra extrinsic scatter and the severe overlap between the red and
blue components.

\subsection{Colour Assignment Schemes}
\label{subsec:scheme}

With the mock colours generated from $P^{\mathrm{red}}$ and $P^{\mathrm{blue}}$, we now proceed to construct
three more comprehensive halo quenching mock catalogues for our analysis in \S~\ref{sec:results}.

\subsubsection{Baseline Halo Quenching}
\label{subsubsec:baseline}

The {\it baseline} halo quenching mock catalogue corresponds to our null hypothesis --- the scatter
within either colour population is independent of halo properties, despite the fact that the blue-to-red
transformation is statistically driven by halo mass. In this case, the relative colour of a red galaxy
with respect to the RS ridge-line is independent of its host halo properties, and the galaxies redder
than the ridge-line~(hereafter referred to as the $\mr$ galaxies) would live in similar halos as
the less red galaxies~(hereafter referred to as $\lr$).  Likewise in the blue population, the bluer
half~($\mb$) would mix well with the less blue half~($\lb$) in terms of their dark matter habitats. To
build such a catalogue, we simply assign the red $g{-}r$ colours randomly to the red mock galaxies
by drawing from Equation~\eqref{eqn:pred}~\citep[similar to the `AbM' model in][]{saito16}, and likewise for the blue colours for blue galaxies using
Equation~\eqref{eqn:pblue}. This baseline halo quenching mock should exhibit the minimum level of the
environmental dependence of colours and colour conformity around centrals among all the halo quenching mocks.

\subsubsection{Fiducial Halo Quenching}
\label{subsubsec:fiducial}

Within the halo quenching model, there are at least two channels through which the environmental dependence
of colours could be boosted beyond the baseline mock.  One possibility is that halo mass remains the key
quantity in ``tinting'' the colour of a red or blue galaxy, so that the $\mb$, $\lb$, $\lr$, and $\mr$
galaxies at fixed $\ms$ live in progressively more massive halos. This naturally extends the simple
halo quenching model from modelling binary to continuous colour variables. To build such an extension,
we introduce the cross-correlation coefficients between colour and halo mass at fixed stellar mass,
$\rho^{\mathrm{cen}}_m$ and $\rho^{\mathrm{sat}}_m$, as our two new parameters. As the red fraction of
central galaxies is a steeper function of halo mass than that of satellites, we choose a higher value for
$\rho^{\mathrm{cen}}_m$~(0.5) than $\rho^{\mathrm{sat}}_m$~(0.3) for galaxies with $\lg\ms\,{<}\,11$;
For the reasons outlined in \S~\ref{subsec:scatter}, at $\lg\ms\,{\geqslant}\,11$ we reduce the value
of $\rho^{\mathrm{sat}}_m$ to zero, while keeping the value of $\rho^{\mathrm{cen}}_m$ at $0.5$ --- we
assume that the increase in the extrinsic scatter of colour is mainly due to the enhanced activity level
of satellites in massive halos.  We hereafter adopt this catalogue as our {\it fiducial} halo quenching mock.

\subsubsection{``Assembly-Biased'' Halo Quenching}
\label{subsubsec:cass}

Alternatively, galaxy assembly bias could coherently modulate the quenching processes within the same
large-scale environment, making galaxies slightly redder~(bluer) in over~(under)-dense regions than in the
field. If such galaxy assembly bias effect is mediated via some secondary halo property other than $\mh$, we
would expect a strong correlation between galaxy colour and that mediator. For example, one might speculate
that the older halos host slightly redder galaxies than the younger ones at the same mass, if they have had
more time forming and quenching galaxies in denser environments. However, the theoretical connection between
any secondary halo properties and galaxy formation remains obscure. But for the purpose of our study, it is
not necessary to distinguish which halo property is the true underlying mediator, as long as it is a good
indicator for halo assembly bias that strongly correlates with large-scale overdensity.

Halo assembly bias reveals itself as the dependence of halo clustering on a variety of
secondary halo properties, the three most prominent of which are concentration, age, and
spin~\citep{sheth04,wechsler06,gao07,lee2017}. Among the three, the age dependence is much weaker than
the other two on the high mass end~\citep{jing07}, while the impact of halo spin on galaxy formation
is likely the most complicated~\citep{vandenbosch02, lacerna12, dubois2014, teklu15}, depending on
spin-tidal field alignment~\citep{aragon-calvo07, zhang09, libeskind13, shi15, kang15} and halo merger
history~\citep{donghia07, welker14, bett16, gomez16}. Therefore, we will focus on halo concentration as
our proxy for galaxy assembly bias.

\begin{figure}
\begin{center}
    \includegraphics[width=0.4\textwidth]{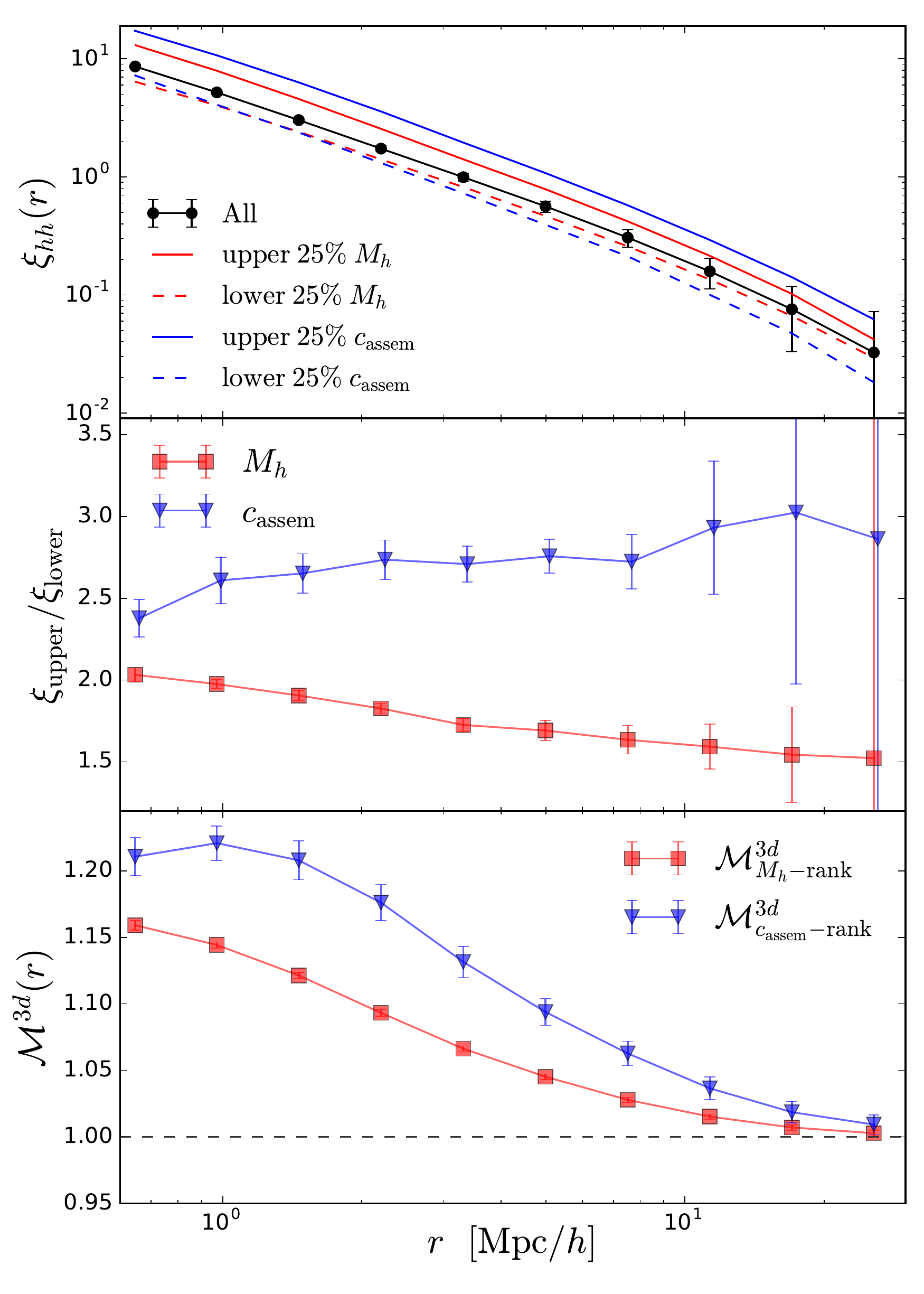} \caption[]{{\it Top panel}:
    3D real-space correlation functions~($\xi_{hh}$) of dark matter halos in the \bolshoi\ simulation.
    Black circles with errorbars show the measurement for all halos with $\mh{>}10^{11}\hmsol$. Red solid
    and dashed curves are the $\xi_{hh}$ for halos in the upper and lower $25$ per cent tails of the $\mh$
    distribution, while the blue solid and dashed curves indicate the upper and lower quartiles of
    $\cass$. As defined by Equation~\ref{eqn:cass}, $\cass$
is a proxy for the halo assembly bias effect, and the clustering difference between the two $\cass$ quartiles
is thus solely driven by halo assembly bias, independent of any halo-mass-related trends~(see
    text for details).	{\it Middle panel}: Ratios between the $\xi_{hh}$ of halos in the upper and lower
    quartiles of $\mh$~(red squares) and $\cass$~(blue triangles), respectively. {\it Bottom panel}:
    Mark correlation functions of halos with $\mh{>}10^{11}\hmsol$, using $\mh$-rank~(red squares)
    and $\cass$-rank~(blue triangles) as marks, respectively.}
\label{fig:halomass_conformity2}
\end{center}
\end{figure}

\begin{figure*}
\begin{center}
    \includegraphics[width=1.\textwidth]{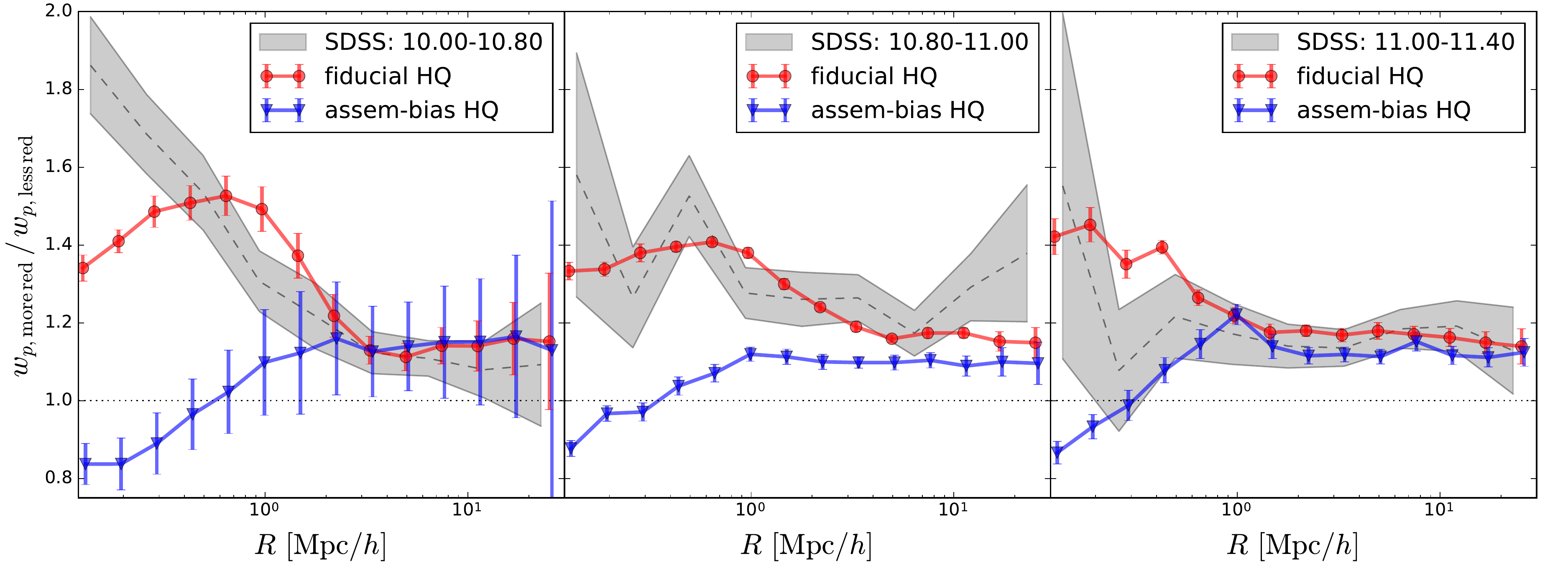} \caption[]{Ratio between the
    projected cross-correlation functions $w_p$ of more red vs.\ less red galaxies (both cross-correlated
    with the parent red galaxy sample) within three different stellar mass bins. In each panel, the
    dashed curve with the gray uncertainty band is measured from a volume-limited sample in SDSS, while
    red circles and blue triangles are predictions from our fiducial halo quenching mock and a modified
    halo quenching model with assembly-bias effects, respectively.} \label{fig:wpratioredfinecolor}
\end{center}
\end{figure*}

To remove the anti-correlation between halo concentration and mass~\citep{wechsler06}, we define a new variant
of the concentration parameter for each halo with mass $\mh$,
\begin{eqnarray}
    \ln c_{\mathrm{assem}}&=&\left\{
\begin{array}{lll}
  {\ln c} - {\langle \ln c(\mh) \rangle} &\,\mbox{if $\mh \leqslant M^{nl}$ }\\
  -\big( {\ln c} - \langle \ln c(\mh) \rangle\big) &\,\mbox{if $\mh > M^{nl}$ } ,
\end{array} \right.
\label{eqn:cass}
\end{eqnarray}
where $c$ is the standard halo concentration parameter, $\langle \ln c(\mh) \rangle$ is the average
logarithmic concentration of all halos at $\mh$, and $M^{\mathrm{nl}}$ is the characteristic non-linear mass
scale~($\lg M^{nl}\,{=}\,12.8$ in our mocks). Since halo concentration follows a log-normal distribution at
fixed halo mass, and the logarithmic scatter is roughly constant with halo mass, there is little residual
correlation between $\cass$ and $\mh$. In addition, we flip the sign inside the parentheses at
$M^{\mathrm{nl}}$, because the cross-correlation coefficient between halo concentration and overdensity goes
from being positive to negative across $M^{\mathrm{nl}}$~\citep{gao07, jing07}. Therefore, this new parameter
$c_{\mathrm{assem}}$ serves as our proxy for halo assembly bias, and is positively correlated
with large-scale overdensity on all mass scales.

Similar to the fiducial halo quenching mock, we introduce two new parameters, $\rho^{\mathrm{cen}}_c$ and
$\rho^{\mathrm{sat}}_c$, as the cross-correlation coefficients between colour and $c_{\mathrm{assem}}$ for the
centrals and satellites, respectively. For our {\it assembly-biased} halo quenching catalogue, we set both
cross-correlation coefficients to be unity, thereby maximizing the level of conformity that can be achieved by
turning on galaxy assembly bias within halo quenching.

We emphasize that our assembly-biased halo quenching mock is fundamentally different from a pure assembly-bias
quenching mock like, e.g., the age-matching mock of \citet{hearin14}. At fixed $\ms$, in the assembly-biased
halo quenching mock, the blue-to-red transition is statistically determined by the simple halo quenching model
of Paper II, while the more vs. less red~(blue) colours are driven by their halo assembly bias via
$\cass$; In the age-matching mock, however, the colour of a galaxy at fixed $\ms$ depends almost exclusively
on its halo assembly bias via a characteristic redshift $z_{\mathrm{starve}}$. For most of the centrals~(below
$10^{11}\hhmsol$), $z_{\mathrm{starve}}$ is equivalent to the formation redshift of the subhalos, but at very
high $\ms$ it largely corresponds to the first epoch at which halo mass exceeds $10^{12}\hmsol$.%}

\subsection{Theoretical Sources of Conformity in Mock Catalogues}
\label{subsec:source}

\begin{figure}
\begin{center}
    \includegraphics[width=0.4\textwidth]{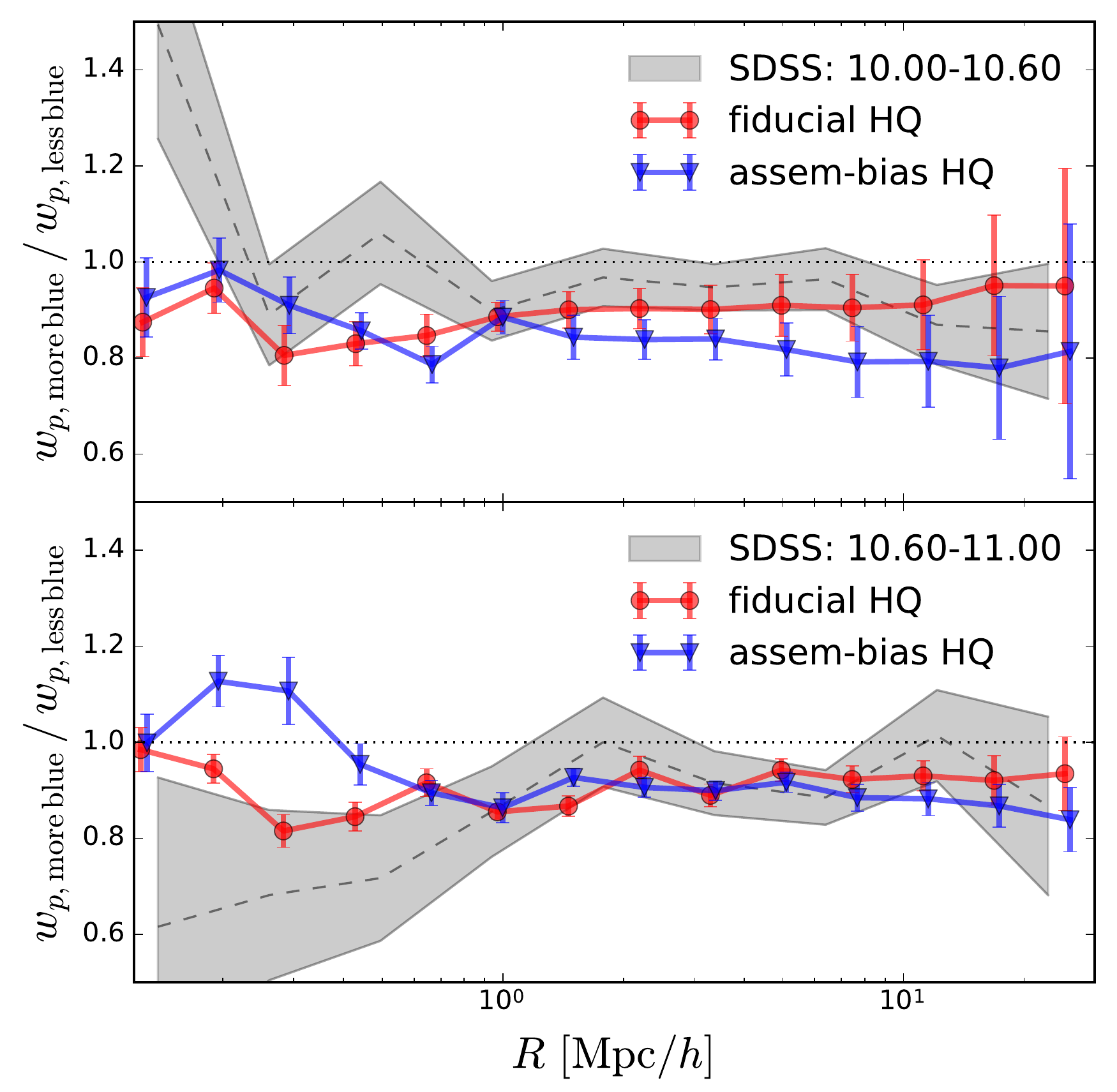}
    \caption[]{Similar to Figure~\ref{fig:wpratioredfinecolor} but for
    more blue vs.\ less blue galaxies within two different stellar mass bins.
}
    \label{fig:wpratiobluefinecolor}
\end{center}
\end{figure}

Before delving into our main results in the next section, we explore the two theoretical sources of conformity
in our three halo quenching mock catalogues in Figure~\ref{fig:halomass_conformity2}. In the top panel, we
show the 3D real-space correlation functions~($\xi_{hh}$) of all \bolshoi\ halos above $10^{11}\hmsol$~(black
circles), along with that of halos in the upper~(solid) and lower~(dashed) quartiles of the distributions in
$\mh$~(red) and $\cass$~(blue). The middle panel shows the ratios between the $\xi_{hh}$ of two halo
subsamples selected by $\mh$~(red squares) and $\cass$~(blue triangles), respectively. As expected from the
standard theory of halo biasing~\citep{sheth99}, the high mass halos have a stronger clustering strength than
the low mass ones on all scales, while the halo assembly bias effect is illustrated by the high vs.
low-$\cass$ halos --- the clustering biases of these two halo subsamples are different despite having the same
average mass~($\langle\lg\mh\rangle{\simeq}11.45$). However, the bias ratios here do not give us full
information on the potential conformity signal that can be induced by $\mh$ or $\cass$.

On the contrary, the mark correlation function is an ideal tool for revealing the environmental dependence of
halo/galaxy properties. The 3D
real-space mark correlation is measured as
\begin{equation}
    \mathcal{M}^{3d}(r) = \frac{WW}{DD} = \frac{1+W(r)}{1+\xi(r)},
\end{equation}
where $WW$ and $DD$ are the mark-weighted and unweighted number counts of pairs with 3D distance
$r$~\citep{sheth05, skibba06}, while $W(r)$ and $\xi(r)$ are the mark-weighted and unweighted real-space
correlation functions, respectively.  Since the marks are normalized to have a mean of unity, if the
distribution of marks is independent of environment, $\mathcal{M}^{3d}(r)$ should be unity on all scales due to the
lack of conformity. Otherwise, if the marks of two objects are correlated over some scale, $\mathcal{M}^{3d}(r)$
will deviate above unity at that scale.

The bottom panel of Figure~\ref{fig:halomass_conformity2} shows the 3D real-space mark correlation functions
of halos, using the rank-orders of $\mh$~(red squares) and $\cass$~(blue triangles) as marks. We use
rank-orders so that the distributions of the two marks are the same uniform distribution from $0$ to $2$,
making a direct comparison between the two mark correlations feasible~\citep{skibba13}.  Both mark correlation
signals decline with radius, but stay significantly above unity on scales up to $15\,\hmpc$, exhibiting strong
conformities in the 2-halo regime. Therefore, we expect that an environmental dependence of colours~(including
a 2-halo colour conformity) would naturally arise in all three halo quenching mocks, but the overall amplitude
and scale-dependence would be different from one to another, due to the varying relative strength of the mass
effect and assembly bias in each mock.

Finally, we emphasize that the underlying drivers of the environmental effects in galaxy colours are
different among the three mocks --- while the baseline and the fiducial halo quenching mocks rely solely on
the environmental dependence of the halo mass function to produce an ``indirect'' environmental effect, the
assembly-biased halo quenching mock predicts both direct and indirect environmental effects, via the halo
assembly bias effect of $\cass$ and the environmental dependence of halo mass functions, respectively.
Meanwhile, the age-matching mock employs the halo assembly bias effect of $z_{\mathrm{starve}}$ to produce
direct environmental effects in the spatial distribution of galaxy colours.

\section{Results}
\label{sec:results}

In this Section, we present the main results of our analysis, by comparing the three halo quenching mock
catalogues with the SDSS data using different measurements. These include the clustering $w_p$ and weak lensing
$\ds$ of the $\mr$ and $\lr$ galaxies, the 2D mark correlation functions of colours $\mk$, and the red galaxy
fractions around red vs.\ blue primaries $\fred$.

\subsection{Clustering and Lensing of ``More Red'' vs.\ ``Less Red'' Galaxies}
\label{subsec:sanity}

\begin{figure*}
\begin{center}
    \includegraphics[width=1.\textwidth]{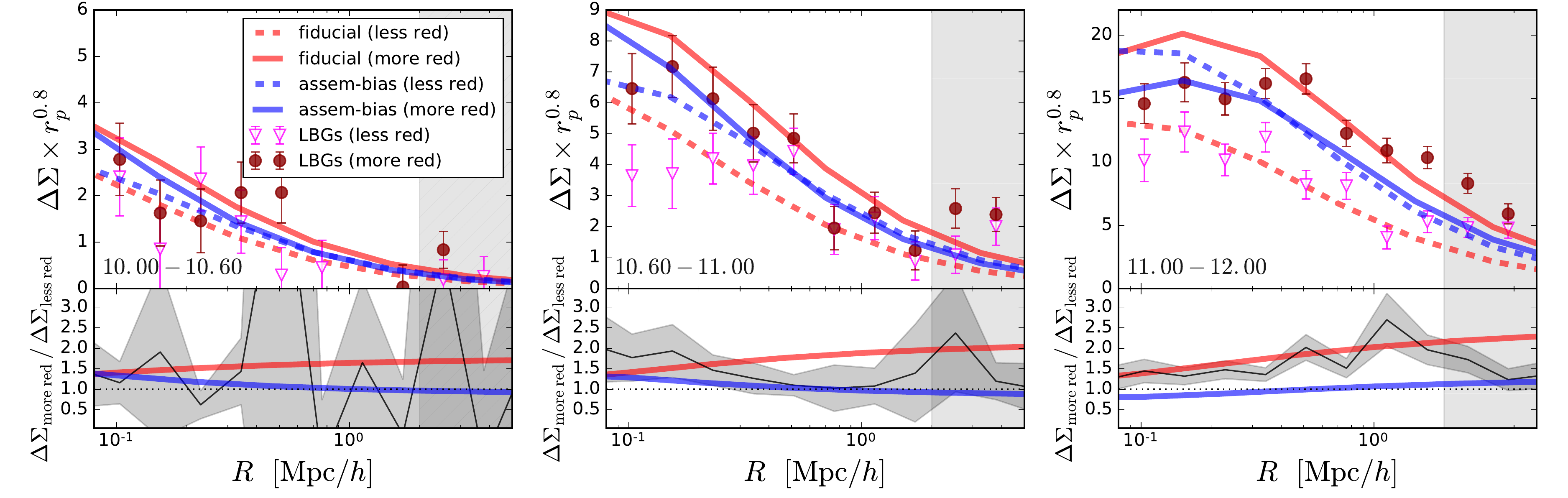} \caption[]{{\it Top panels:}
    Comparison between weak lensing signals
	measured from more red vs.\ less red locally brightest galaxy samples
    derived from SDSS~(red filled circles and magenta open triangles with errorbars), and those predicted
    by our fiducial halo
	quenching mock~(red solid and dashed curves) and an assembly-bias modified halo quenching
	mock~(blue solid and dashed curves), for three
    different stellar mass bins. The model curves (predictd using only the 1-halo term) cannot be directly
    compared with data inside the gray vertical bands, which indicate the scales at which the 2-halo
    term becomes important. {\it Bottom panels:} Ratios between the $\ds$ profiles of the more red vs.\
    less red LBGs, as measured from weak lensing in SDSS~(black curves with gray uncertainty bands),
    and predicted from the fiducial~(red) and assembly-bias modified~(blue) halo quenching models.}
    \label{fig:fineredcolor_lbgwl}
\end{center}
\end{figure*}

As a sanity check, we will first compare the clustering and g-g lensing of the $\mr$~(more red) and
$\lr$~(less red) galaxies predicted by the mocks to that measured from data. By construction, the three halo
quenching mocks assign different values of red $g{-}r$ colours to the red galaxies that share the same set of
halos in the simulation, therefore placing the $\mr$ and $\lr$ galaxies into different halos.  In particular,
the $\mr$ and $\lr$ galaxies occupy halos with similar $\mh$ and $\cass$ in the baseline mock, but live in
halos of different $\mh$ or $\cass$ in the other two mocks. This segregation of galaxies by halo properties,
which is lacking in the baseline mock, dictates that the $\mr$ and $\lr$ galaxies will exhibit different
clustering and weak lensing signals, a prediction that can be directly tested using the SDSS data.

Figure~\ref{fig:wpratioredfinecolor} shows the ratios of the projected correlation functions $w_p$ between
the $\mr$ and $\lr$ galaxies for three different stellar mass bins. %{\bf
To obtain a more stable measurements of the clustering ratios, we compute the cross-correlations of the
two subsamples with the overall red galaxy sample in each bin, instead of the auto-correlations shown
in Figure~\ref{fig:wpcolor}.  In each panel, the gray dashed curve with shaded band is
the measurement from volume-limited samples in SDSS, and the data points with errorbars are predictions
by the fiducial~(red circles) and assembly-biased~(blue triangles) halo quenching mocks. Clearly,
the baseline mock, which predicts a $w_p$ ratio of unity on all scales~(dotted horizontal line),
is readily ruled out in all three stellar mass bins. The fiducial halo quenching mock shows the best
overall agreement with the SDSS measurements on all scales between $0.1$ and $25\,\hmpc$,
 whereas the assembly-biased halo quenching mock only provides an adequate
description of the SDSS measurements on scales beyond $2\,\hmpc$~(except for the intermediate mass bin
where it fails to match the data on all scales). It is worth nothing that some of the discrepancies
between the mock predictions and the SDSS measurements on small scales are caused by the lack of colour
segregation in the mocks~(e.g., the rapid increase of observed ratios on scales below $0.2\,\hmpc$ in SDSS).

The assembly-biased halo quenching mock predicts that the $\mr$ and $\lr$ galaxies live in halos of
similar mass~(mostly above $M^{nl}$), hence the similar small-scale amplitude of $w_p$. Furthermore,
it predicts that the $\mr$ satellite galaxies live in high-$\cass$ halos, which mainly correspond to
low-$c$ halos and a less concentrated satellite distribution within halos, causing the ratio to cross
from above to below unity on scales ${\lesssim}\,0.5\,\hmpc$.  In the low~($\lg\ms\,{=}\,[10.0,\,10.8]$)
and intermediate~($\lg\ms\,{=}\,[10.8,\,11.0]$) stellar mass bins, however, neither the decrease of the
ratio on small scales nor the ratio inversion at $0.5\,\hmpc$ is seen in the data, indicating that the
assembly-biased halo quenching is not an adequate model for the colouring of red-sequence {\it satellite} galaxies
below $10^{11}\,\hhmsol$.
For the high mass bin~($\lg\ms\,{=}\,[11.0,\,11.4]$), the SDSS measurement
is consistent with predictions from both mocks, in part due to the large errorbars resulting in poor
discriminating power on the small scales where the predictions from those mocks differ.

 Similarly, Figure~\ref{fig:wpratiobluefinecolor} shows the ratio comparison between the more blue vs.\ less
blue galaxies in two different stellar mass bins. Due to the lack of a distinct colour ridgeline in the
blue portion of the colour-stellar mass diagram, we divide each blue sample into two halves using the
median blue colour as a function of $\ms$. At any given stellar mass, the two subsamples of blue galaxies
in SDSS exhibit weaker discrepancies in their clustering biases than the two red galaxy subsamples,
and the
observed clustering ratio is also better-reproduced by the blue galaxies in the fiducial mock than
those in the assembly-biased halo quenching mock on all scales.

For diagnosing the colour properties of central galaxies, we turn to the weak lensing signals around central
galaxies with different shades of red. Following \citet{mandelbaum16}, we select a sample of ``locally
brightest galaxies''~\citep[LBGs;][]{planck13} from the SDSS spectroscopic sample as our candidates for
central galaxies, and measure the weak lensing profiles for the red LBG subsamples split by the RS ridgeline
at fixed stellar mass.  \citet{mandelbaum16} discovered a strong bimodality in the average halo mass between
the red and blue LBGs, which we subsequently interpreted as the indication of a dominant halo quenching
mechanism using our \ihod\ framework in Paper II.

In the same spirit as \citet{mandelbaum16}, we show the weak lensing profiles $\ds$ of the $\mr$~(maroon
filled circles) and $\lr$~(magenta open triangles) LBGs within three stellar mass bins in the upper
panels of Figure~\ref{fig:fineredcolor_lbgwl}. In each upper panel, we also show the predictions from the
fiducial~(red solid and dashed curves) and assembly-biased~(blue solid and dashed) halo quenching mocks.
We ignore the profiles beyond $R\,{=}\,2\hmpc$~(covered by the gray shaded region) as they do not carry
clean information on the halo mass profile. In the lower panels, we show the ratios between the $\mr$
and $\lr$ weak lensing profiles for the SDSS LBGs~(black curve with gray shaded bands), the fiducial
halo quenching mock~(red curve), and the assembly-biased halo quenching mock~(blue curve). As expected,
the assembly-biased halo quenching mock predicts a $\ds$ ratio of roughly unity on all scales, modulo
minor tilt due to differences in halo concentration. But in the fiducial mock, the $\mr$ central galaxies
exhibit stronger weak lensing amplitude than the $\lr$ centrals on the relevant scales.

In the lowest mass bin~($\lg\ms\,{=}\,[10.0,\,10.6]$), both mock predictions are consistent with the data, but
the large uncertainties in the weak lensing measurements prevent us from making any statistical
statements on one mock being preferred by the data. Similarly in
the intermediate mass bin~($\lg\ms\,{=}\,[10.6,\,11.0]$), both mock predictions are roughly consistent with
the measured $\ds$ profiles. However, the $\ds$ ratio measurement on scales above $0.6\,\hmpc$ slightly
prefers the assembly biased halo quenching mock, but shows a strong discrepancy between the two subsamples on
scales below $0.6\,\hmpc$, which tends to favor the fiducial halo quenching mock. Fortunately, the weak
lensing measurements in the high mass bin~($\lg\ms\,{=}\,[11.0,\,12.0]$) leave no ambiguity as to which mock
is the superior model for colouring high-mass central galaxies --- the fiducial halo quenching mock provides
an excellent description for the $\ds$ profiles of the more vs.\ less red LBGs and the ratio between the two, while
the assembly-biased halo quench mock completely fails to do so.

Combining Figures~\ref{fig:wpratioredfinecolor},~\ref{fig:wpratiobluefinecolor}, and~\ref{fig:fineredcolor_lbgwl}, it is clear that the fiducial
halo quenching mock significantly out-performs the other two mocks in describing the clustering and weak
lensing of red galaxies split by the RS ridge-line. However, the weak lensing measurements for low-$\ms$
galaxies have large uncertainties, and we cannot perform the same lensing test on the blue galaxies due to the
their low number density in this mass range. With this result and its limitations in mind, we will
turn to the mark correlation functions of colours for a clearer picture in the next subsection.

\subsection{Environmental Dependence: Mark Correlation Functions of Galaxy Colours}
\label{subsec:mark}

\begin{figure*}
\begin{center}
    \includegraphics[width=1.\textwidth]{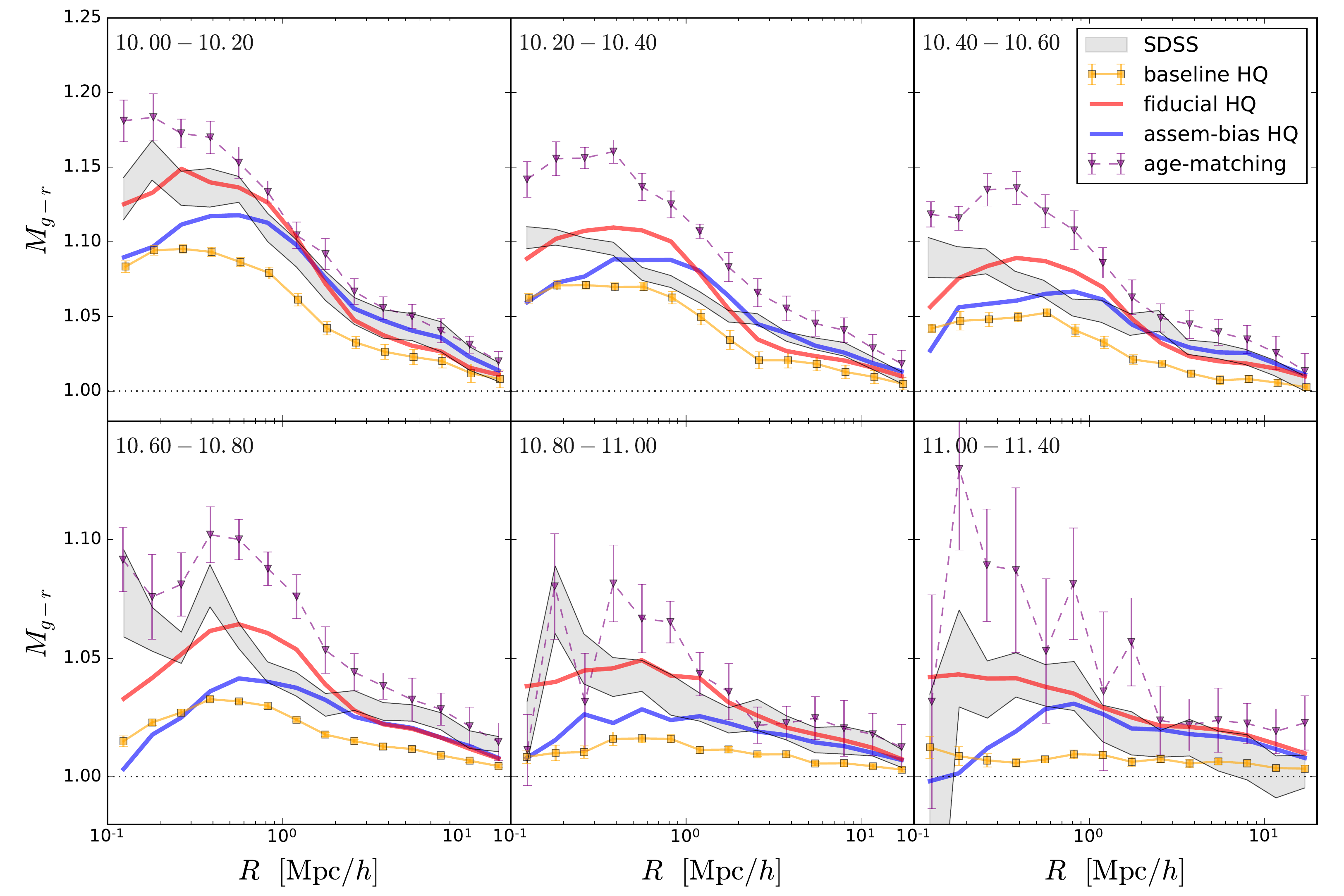} \caption[]{Mark correlation functions of
    galaxy $g{-}r$ colours within six different stellar mass bins. In
each panel, the gray uncertainty band indicates the mark correlation measured from a volume-limited galaxy
sample in SDSS. Orange squares with errorbars show the prediction from the baseline halo quenching mock,
while red and blue curves are predictions from the fiducial and assembly-bias modified halo quenching mocks.
The sizes of errorbars~(not shown here) on the red and blue curves are comparable to that in the baseline
model. Purple triangles with errorbars are predicted from an age-matching mock with strong assembly bias
effect on colours~\citep{hearin14}. Among the four models, the fiducial halo quenching mock provides the best
overall description for the scale and stellar mass dependence of colours in SDSS.}
  \label{fig:mk}
\end{center}
\end{figure*}

Marked statistics are an efficient tool for quantifying the correlation between the properties~(i.e.,
marks) of galaxies and their environment~\citep{martinez10}. Here we focus on the 2D projected mark
correlation function of $g{-}r$ colours $M_{g{-}r}(R)$ as our diagnostic of the underlying driver of
environmental dependence of galaxy colours, whether it be halo mass, halo assembly bias, or some combination
of the two.

Following earlier studies~\citep{skibba06, skibba09, skibba12} of galaxy mark correlation functions in SDSS,
we define $M_{g{-}r}(R)$ as
\begin{equation}
    M_{g{-}r}(R) = \frac{1+W_p(R)/R}{1+w_p(R)/R},
    \label{eqn:m2d}
\end{equation}
where
$W_p$ and $w_p$ are the mark-weighted and unweighted projected correlation
functions of galaxies, respectively. This particular definition makes $M(R)\,{\sim}\,\mathcal{M}^{3d}(r{\equiv}2R)$
on large scales, where $w_p(R)/R\,{\sim}\,\xi$.
% $w_p(R)/R\,{\simeq}\,\xi(aR)$ for a power-law $\xi\,\propto\,r^{-\gamma}$. For a
% typical galaxy clustering with $\gamma\,{\sim}\,1.8$, $a$ is approximately $2$.

We measure $M_{g{-}r}(R)$ by combining the measurements of $W_p$ and $w_p$ as in Equation~\ref{eqn:m2d}, and
compute the uncertainties by Jackknife re-sampling $200$ sub-regions within the sample footprint. We refer
readers to Paper I for the technical details on the projected correlation function measurements. To make sure
that the mark values are always positive, we adopt $\exp(g{-}r)/\langle\exp(g{-}r)\rangle$ as the mark in both
the data and mock catalogues. Since the mock colour distribution at each fixed $\ms$ is almost identical to
that in SDSS, we can compare the colour-mark correlation functions of the mock and data catalogues directly.
More importantly, since all the mocks accurately reproduce the abundance, spatial clustering, and g-g lensing
of the red and blue galaxies in SDSS, any discrepancy between the colour-mark correlations of the mock and the
data galaxies would be a clean sign of incorrect colour assignment in that mock.

Figure~\ref{fig:mk} compares the colour-mark correlation functions between the data and various mocks within
six different stellar mass bins. In each panel, the gray shaded band indicates the measurement~(${\pm}\,1\sigma$
range) from a volume-limited sample of SDSS galaxies within the $\lg\ms$ range~(listed on the top left);
Orange
squares with errorbars are measured from the baseline halo quenching mock, while the red and blue thick curves
are from the fiducial and assembly-biased halo quenching mocks, respectively, with similar uncertainties~(not
shown) as in the baseline case. We also show the measurement~(magenta triangles with errorbars) from the
age-matching mock produced by
\citet{hearin14}~\footnote{\url{http://logrus.uchicago.edu/~aphearin/SDSS_Mock_Catalog/SDSS_Mock_Catalog.html}}, which serves as an interesting limiting case in which the
galaxy assembly bias effect is maximized and the halo quenching effect is minimized. Overall, the observed
mark correlation signal decreases with increasing stellar mass, partly due the lack of a prominent blue
population in the high-$\ms$ samples. The observed $M_{g{-}r}$ signal also decreases as a function of
distance, analogous to the 1-halo to 2-halo transition in the regular correlation functions. However, in all
the stellar mass bins below $10^{11}\,\hhmsol$, the observed mark correlations stay significantly above unity
on scales up to $15\,\hmpc$, indicating strong environmental dependence of galaxy colours at
fixed $\ms$.

The colour-mark correlation functions predicted by the baseline halo quenching model~(orange squares with
errorbars) have similar shapes as the $M_{g{-}r}$ measured from SDSS, but their overall amplitudes
are $30$-$50\%$ lower than the observed ones.  The lower amplitudes are expected: by design the baseline
model includes the least amount of halo quenching effect that is allowed by the measurements of the
clustering and g-g lensing of red and blue galaxies in SDSS.

With a stronger coupling between halo mass and galaxy colour, the fiducial halo quenching mocks~(red curves)
are able to roughly reproduce the correct amplitudes of $M_{g{-}r}$, especially the deviation from unity on
large scales.
Similar to the baseline model, the fiducial halo quenching
mock is consistent with the clustering and lensing measurements of the red and blue galaxies in SDSS. In
addition, the fiducial model is also the only one among the three colour assignment schemes that is consistent
with the observations of the clustering and lensing of the more vs.\ less red~(blue) galaxies. Therefore, it is
tempting based on the combined evidence so far to argue that the fiducial halo quenching mock roughly
encodes the correct information on the connection between galaxy quenching and halo properties in SDSS. The
primary
discrepancy between the mark correlations predicted by the fiducial model and the data is that the mock
curves have a slightly sharper transition from 1-halo to 2-halo scales compared to the data, probably due to
the lack of colour segregation and the sharp halo boundaries adopted in the mock.  It is worth
noting that the 1-halo to 2-halo transition is notoriously difficult to model correctly even for
regular correlation functions (not marked).

The assembly-biased halo quenching mock predicts a very similar amplitude of $M_{g{-}r}$ on large scales as
the fiducial mock and the data, and is thus consistent with the SDSS measurements. However, recall that the
assembly-biased model represents the maximum amount of galaxy assembly bias~(mediated by concentration) that
can be allowed in the halo quenching scenario --- the colour of a red/blue galaxies is maximally coupled with
the rank-order of $\cass$, our proxy for the halo environment; therefore, it is very unlikely that halo
concentration is the underlying driver for the ``tinting'' of red or blue galaxies under the halo quenching
framework.  On small scales, the predicted curves level off and approach unity for the three stellar mass bins
above $\lg\ms{=}10.6$. Similar to the small-scale behavior seen in Figure~\ref{fig:wpratioredfinecolor}~(blue
curves), the low amplitudes on scales below $0.3\,\hkpc$ are caused by the lower values of $g{-}r$ colours
assigned to the satellites in massive halos with high concentrations than those inside low-concentration
clusters.

Unlike the three halo quenching mocks, the age-matching mock relies on galaxy assembly bias, i.e., by matching
galaxy colours to their halo formation times at fixed $\ms$, to introduce the environmental dependence of
galaxy colours. Clearly, the maximal coupling between galaxy colour and halo age is strongly
disfavored by the data, as the predicted mark correlation functions~(magenta triangles with
errorbars) are $40$-$50\%$ higher than the observations on almost all scales and for four of the six stellar
mass bins. A more reasonable fit to the data can be potentially achieved by tuning down the coupling strength
between galaxy colour and halo age, but as we pointed out in Sec.~\ref{sec:intro}, without halo quenching it is
still unlikely to reproduce the observed bimodality in the host halo mass of red vs.\ blue central
galaxies.  We need to globally match not just the mark correlation functions, but also the regular
two-point correlations, for which Paper I and~II demonstrated halo quenching is necessary.

\begin{figure*}
\begin{center}
    \includegraphics[width=1.\textwidth]{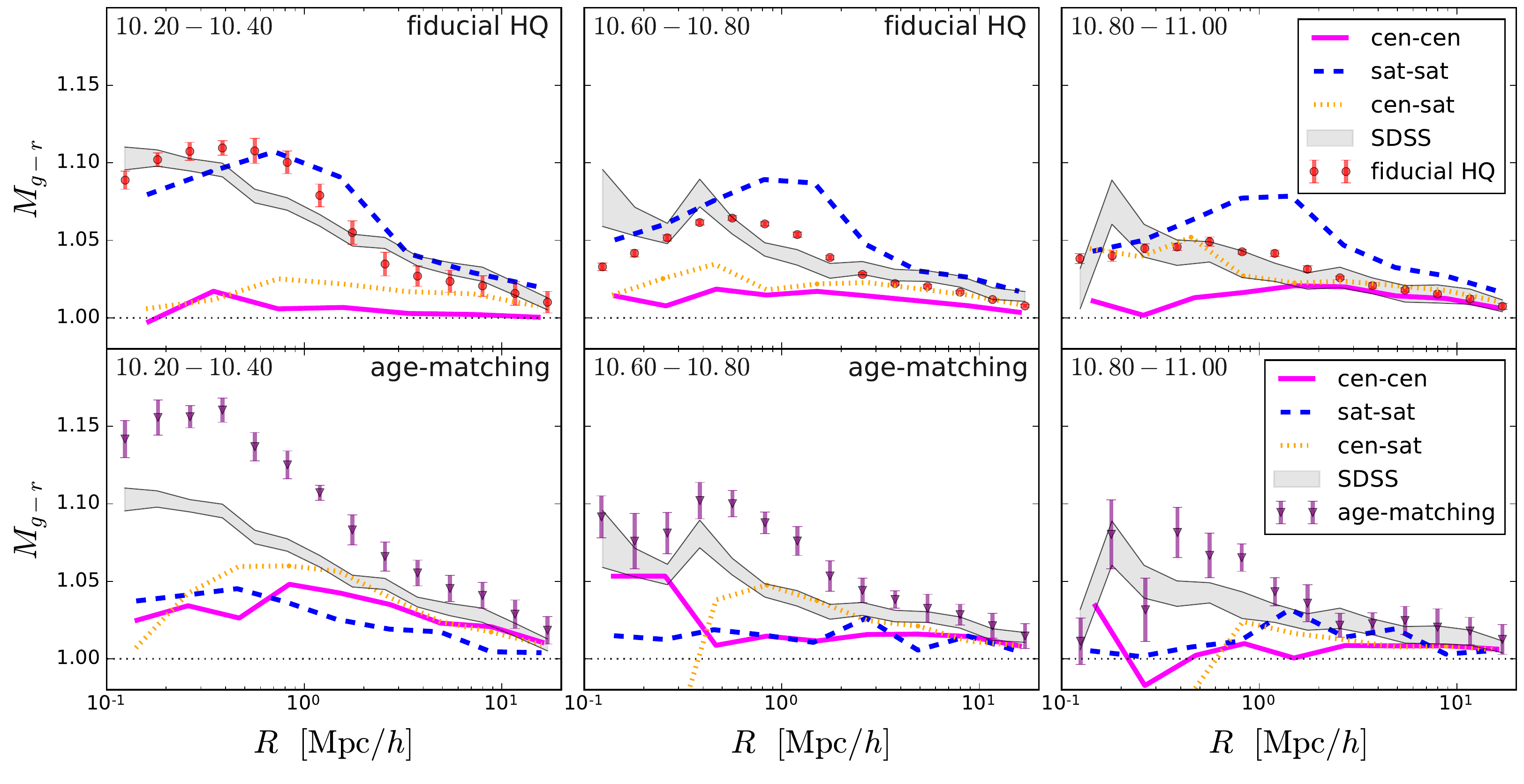} \caption[]{Mark correlation functions
    of \texttt{cen-cen}~(solid magenta), \texttt{cen-sat}~(dotted
	orange), and \texttt{sat-sat}~(dashed blue) pairs, predicted by the fiducial halo quenching~(upper
	panels) and the age-matching~(lower panels) mocks for three different stellar mass bins. In each
	panel, we also show the prediction for the overall mark correlation function~(symbols with errorbars)
	and the SDSS measurement~(gray shaded band). The combination of the \texttt{cen-cen} and
    \texttt{cen-sat} components indicates the level of conformity in each model.}
    \label{fig:mk_components}
\end{center}
\end{figure*}

Clearly, Figure~\ref{fig:mk} shows that there is significant environmental dependence
  of galaxy colours up to scales as large as $15\,\hmpc$, and the fiducial halo quenching model is able to
reproduce this large-scale environmental dependence of colours across the entire stellar mass range we
probed~(${>}\,10^{10}\,\hhmsol$). However, a strong environmental effect revealed by the mark correlation
functions does not necessarily translate to a strong colour conformity, which is exclusively a phenomenon surrounding
central galaxies. In particular, the mark correlation function has two contributing sources: one is the
galactic conformity between the central galaxies and neighbouring galaxies~(i.e., the \texttt{cen-cen} and
\texttt{cen-sat} terms), and the other is the correlation between the colours of satellites~(i.e., the
\texttt{sat-sat} term). Therefore, it is possible that the \texttt{sat-sat} contribution could dominate the
signal in the mark correlation functions, while the conformity contribution stays confined within the 1-halo
regime. If proven true, this scenario would explain the findings by \citet{hearin15}, that an HOD model of
galaxy colours cannot induce 2-halo conformity, despite of the strong environmental effect in the spatial
distribution of galaxy colours.

Figure~\ref{fig:mk_components} compares the three contributions, \texttt{cen-cen}~(magenta solid),
\texttt{cen-sat}~(orange dotted), \texttt{sat-sat}~(blue dashed), between the fiducial halo quenching~(upper
panels) and the age-matching~(lower panels) mocks in three stellar mass bins~(columns from left to right:
$[10.2,\, 10.4]$, $[10.6,\, 10.8]$, $[10.8,\, 11.0]$). In each panel we also show the overall mark correlation
functions measured in the mock~(symbols with errorbars) and the data~(gray shaded bands) from
Figure~\ref{fig:mk}.

In the low-$\ms$ bin~(left column), the environmental dependence of colours in the fiducial halo
quenching mock~(upper left) is primarily driven by the correlation between the satellite colours, and
partly by the colour conformity between central galaxies and satellites. In halo quenching, the dominant
\texttt{sat-sat} term is caused by the strong halo quenching effect on the satellites of rich groups
and clusters, which tend to live in dense environments; The \texttt{cen-cen} term is very weak, as in
this low stellar mass regime the red vs. blue central galaxies are living in halos whose masses differ
only by a factor of two or less, due to the weak halo quenching effect below $10^{12}\,\hmsol$. On the
contrary, the age-quenching mock~(lower left) predicts strong \texttt{cen-cen} and \texttt{cen-sat}
terms, because the host halos of those centrals~(with $\mh\,{<}\,M^{nl}$) exhibit a strong assembly
bias effect~\citep{gao05}; The \texttt{sat-sat} term is substantially weaker than the conformity terms,
due to the weaker halo assembly bias effect in their massive host halos~(with $\mh\,{>}\,M^{nl}$).

At higher stellar massses~(middle and right columns), the \texttt{sat-sat} term remains dominant in the upper
panels~(halo quenching), but the \texttt{cen-cen} term begin to increase, indicating stronger conformity
effects for higher $\ms$ primaries in the halo quenching scenario. For the age-matching model, the
\texttt{cen-cen} term decreases at higher $\ms$, because the halo assembly bias effect becomes weaker with
increasing $\mh$~\citep{jing07}.

From Figure~\ref{fig:mk_components}, we can see that the environmental dependence of colours shown in the
halo quenching mocks are largely driven by the correlation between the colours of satellites. This finding
explains the lack of large-scale conformity in the HOD model of colours in \citet{hearin15} --- a strong
mark correlation function signal does not necessarily imply strong conformity on large scales, especially
for galaxies with $\lg\ms\,{<}\,10.4$.  However, the 2-halo conformity signal predicted by our fiducial
halo quenching mock increases rapidly with stellar mass, becoming comparable to the age-matching prediction
at $\lg\ms\,{\sim}\,10.6$, and much stronger than age-matching at $\lg\ms\,{>}\,10.8$.  Furthermore, it
is worth nothing that even in the lowest stellar mass bin probed here~($\lg\ms\,{\sim}\,10.2$), there
still exists some colour conformity signal on large scales, mostly due to the correlation between the
colours of central galaxies and that of satellites inside nearby massive halos. In the next subsection,
we will explore whether such a level of 2-halo conformity is consistent with the observed red galaxy
fractions around red vs.\ blue primaries in SDSS.

\subsection{Conformity: Red Galaxy Fraction Around Red vs.\ Blue Isolated Primaries}

\begin{figure}
\begin{center}
    \includegraphics[width=0.48\textwidth]{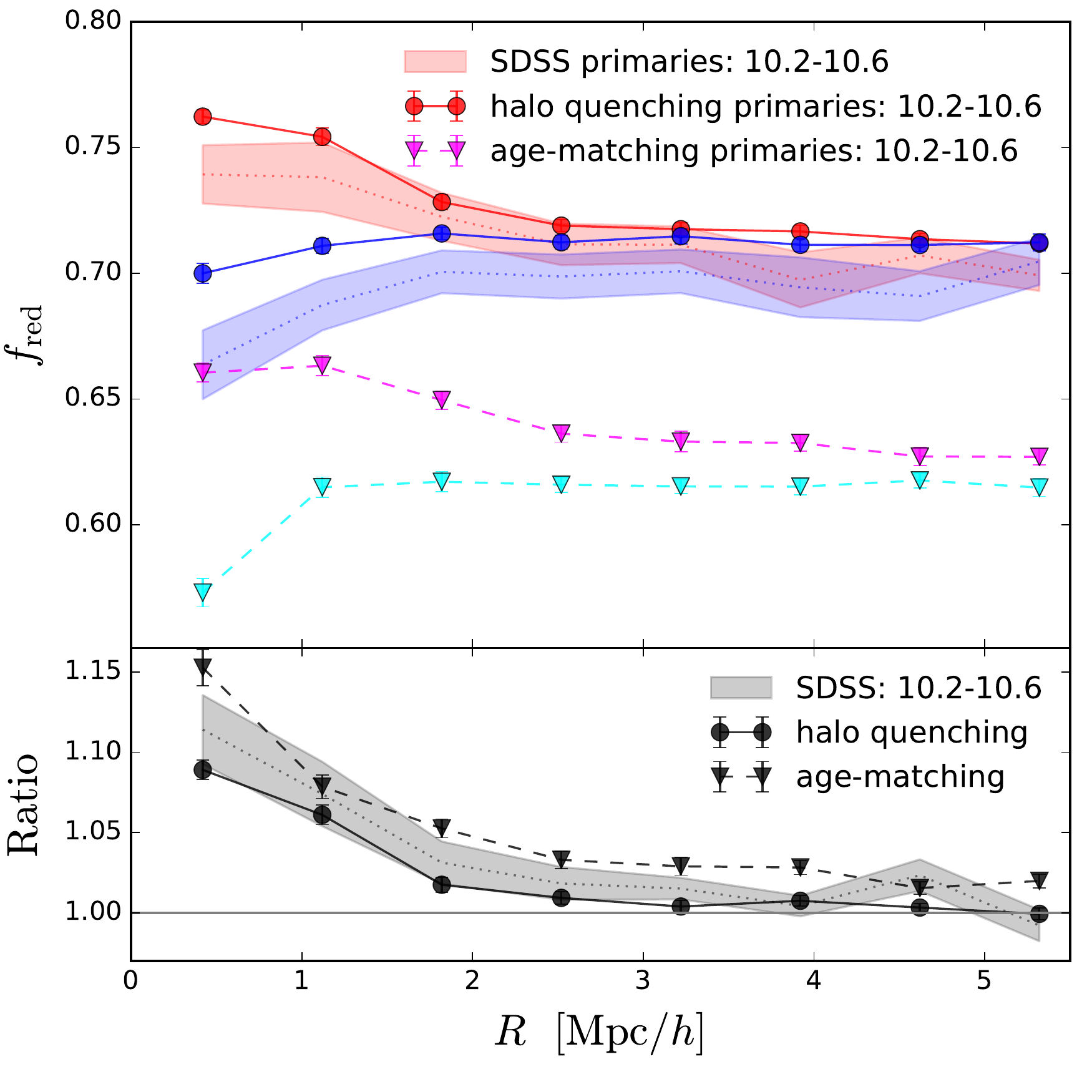} \caption[]{Red galaxy fractions as
    functions of projected distance from red vs.\ blue isolated primaries with $\lg\ms{=}[10.2-10.6]$. In
    the top panel, dotted curves with red and blue uncertainty bands are the measurements from a
    volume-limited sample in SDSS. The red and blue circles with errorbars are the predictions from
    our fiducial halo quenching mock, while the magenta and cyan triangles are from the age-matching
    mock~(the lower average $f_{\mathrm{red}}$ in the age-matching results is due to sample incompleteness
    in the mock). The bottom panels show the ratios between the red fraction profiles of the red vs.\
    blue isolated primaries, for SDSS galaxies~(gray bands), our fiducial halo quenching mock~(circles),
    and the age-matching mock~(triangles). The deviations of ratio profiles from unity indicate colour
    conformity.} \label{fig:conformity}
\end{center}
\end{figure}

So far, the fiducial halo quenching mock has successfully passed the sanity check in \S~\ref{subsec:sanity}
and reproduced the observed strong environmental dependence of colours in~\S~\ref{subsec:mark}.  As we
alluded to in Section~\ref{sec:intro}, \citet{tinker17} and \citet{sin17} showed that the observed level of
large-scale conformity is very sensitive to the isolation criteria for identifying primary galaxies. They
found significant reduction in the conformity signal around low-mass primary galaxies after removing
a small fraction of mis-identified primaries, which are revealed by the group finders~\citep{yang07,
tinker11} to be living inside or near massive systems~(e.g., the Coma cluster). However, based on the
iterative reshift-space friends-of-friends technique, the group finders also have their own limitations
in crowded environments~\citep{campbell15}.  Since our goal here is to find out whether the level of
conformity predicted by the mocks is consistent with observations, the conclusion should not depend on
the criteria for finding centrals, as long as we self-consistently apply the same criteria to the mock
and data catalogues.  Therefore, we will adopt the simple isolation criteria of \citet{kauffmann13},
and directly measure the red galaxy fractions around the red and blue isolated primaries in SDSS, without
resorting to more stringent criteria based on complicated group finders.

For the candidate sample for primary galaxies, we choose the log-stellar mass range of $[10.2,\,10.6]$,
similar to the main stellar mass range probed in \citet{kauffmann13} and \citet{hearin15}. For the secondary
galaxy sample, we select all galaxies with stellar mass above $5\,\times\,10^9\,\hhmsol$, $0.5$ dex below the
minimum stellar mass of the primaries. We adopt the same criteria in \citet{kauffmann13} to find the isolated
primaries --- a galaxy with stellar mass $\ms$ is defined as an isolated primary if there is no other galaxy
with stellar mass greater than $\ms/2$ within a projected radius of $350\,\hkpc$ and with velocity difference
less than $500\,\kms$.

% 0.808147775593
% 0.951342087589

We apply the same sample selection and isolation criteria in the fiducial halo quenching and the age-matching
mock catalogues, and the purities~(i.e., fraction of true central galaxies among the selected isolated
primaries) are $0.81$~(fiducial) and $0.95$~(age-matching), respectively. %{\bf
The lower purity in the \ihod\
    mock is largely due to the higher average frequency of having a satellite galaxy that is at least twice as massive as
the central within halos above $10^{13}\,\hmsol$~(see the lower panels of Fig. 5 in Paper II). %}.
Since the
age-matching mock catalogue is incomplete below $\lg\ms\,{\sim}\,10$, we expect the mean red galaxy fraction
to be different from the observations, but the conformity signal, i.e., the ratio between the red fractions
around red vs. blue primaries, should be preserved. For the observations, we also applied the same isolation
criteria to a volume-limited SDSS galaxy sample with $\ms$ above $5\,{\times}\,10^9\,\hhmsol$ and a redshift
range of $[0.01,\,0.055]$. We then measure the red galaxy fractions as functions of projected distance away
from the red and blue primaries, as well as the ratio between the red galaxy fractions around red vs.\  blue
primaries in each catalogue.

Figure~\ref{fig:conformity} compares the conformity signals measured from the fiducial halo quenching
mock~(circles), the age-matching mock~(triangles), and the SDSS galaxies~(shaded bands), respectively.
All the errorbars are 1-$\sigma$ uncertainties derived from jackknife
re-sampling, and the sizes of the errorbars measured from the two mock catalogues are comparable to the size of
the symbols.
The top panel shows $\fred$, the red galaxy fractions as functions of projected distance away from the isolated
primaries.  The amplitude and scale-dependence of $\fred$ predicted by the fiducial halo quenching mock agree
very well with the observations. This agreement is not by construction, because the \ihod\ quenching
parameters were constrained solely from the clustering and g-g lensing information, without any input from the
relative abundance of red/blue galaxies. The age-matching predictions for the overall red galaxy fractions are
substantially lower, likely due to the sample incompleteness below $10^{10}\,\hhmsol$. However, the age-matching
mock clearly predicts strong conformity, i.e., large discrepancy between the red galaxy fractions around red
vs.\ blue primaries, while the level of conformity in the data and the fiducial halo quenching mock is weaker.

The bottom panel of Figure~\ref{fig:conformity} presents a clearer picture of conformity via the ratio between
the red galaxy fractions around red vs.\ blue primaries. The
SDSS galaxies show substantial colour conformity signal on projected distances up to $2$-$3\,\hmpc$, beyond
which the signal becomes consistent with having no conformity on scales ${\sim}5\,\hmpc$.  The ratio predicted
by the age-matching model is slightly higher than the measurement, and stays significantly above $1.0$ on all
scales. As expected, the ratio predicted by the fiducial halo quenching mock is lower than that predicted by
the age-matching mock on all scales, due to the weaker \texttt{cen-cen} contribution to the environmental
dependence of colours. However, the halo quenching prediction is in good agreement with the SDSS measurement,
showing considerable conformity on scales below $3\,\hmpc$.

Figure~\ref{fig:conformity} demonstrates that the level of conformity observed at stellar mass around a few
${\times}\,10^{10}\,\hhmsol$ can be naturally explained by the HOD model of galaxy colours that includes a
 halo quenching prescription for assignment of colours. Inside the halo quenching mock, the conformity signal is
primarily~($81\%$) sourced by the correlation between the colours of central galaxies and their neighbouring
galaxies, and partly~($19\%$) induced by the mis-identified primaries that are satellites of the more massive
halos. The underlying drivers of both components, however, are the same combination of the environmental
dependence of halo mass function and the strong dependence of galaxy quenching on halo mass --- an indirect
environmental effect imprinted on galaxy formation.

\section{Summary and Discussion}
\label{sec:sum}

\subsection{Fiducial Picture of Halo Quenching}

%{\bf
In this paper, we have investigated whether halo mass quenching is capable of reproducing the environmental
dependence of galaxy colours and the large-scale galactic conformity observed in SDSS, for which recent
studies suggested that a strong galaxy assembly bias effect may be required.  Developed in the first two papers
of the series~\citep{zu15, zu16}, the best-fitting \ihod\ halo quenching model can accurately describe the
observed abundance, spatial clustering, and weak gravitational lensing of the red and blue galaxies in SDSS,
serving as an ideal test-bed for exploring the environmental effects predicted by halo quenching.  For the
prediction, we start by producing quiescent and active mock galaxies within a suite of $N$-body cosmological
simulations based on the best-fitting \ihod\ halo quenching prescription, which describes the red galaxy
fractions of the centrals and satellites as powered-exponential functions of the halo mass. %}
We then assign $g{-}r$ colours to the quiescent and active galaxies separately at fixed stellar mass $\ms$, by
introducing correlations between galaxy colours and the mass of their dark matter halos. In our fiducial halo
quenching model, we set the cross-correlation coefficients $\rho^{\mathrm{cen}}_m$ between halo mass and the
red or blue central galaxies to be $0.5$, while assuming a weaker coupling with halo mass for the red and blue
satellites~($\rho^{\mathrm{cen}}_m\,{=}\,0.3$ below $10^{11}\,\hhmsol$ and $0.0$ above).

Despite the fact that the quantity that determines the galaxy colours in the halo quenching model is
associated with individual halos,
the fiducial halo quenching mock predicts
strong environmental dependence of galaxy colours --- the mark correlation functions of colours deviate
significantly above unity on scales up to $15\,\hmpc$ in all stellar mass bins, in excellent agreement with
the measurements from SDSS. This strong environmental dependence is induced by the combination of halo quenching
and the environmental dependence of the halo mass function: in denser environments there are more massive halos,
hence more quenched and redder galaxies at fixed $\ms$. By decomposing the predicted colour-mark correlation
function at fixed $\ms$ into \texttt{cen-cen}, \texttt{cen-sat}, and \texttt{sat-sat} components, we find that
the overall environmental dependence is dominated by the \texttt{sat-sat} term. However, there still remains
a significant level of \texttt{cen-cen} and \texttt{cen-sat} contributions, i.e., galactic conformity, on
scales up to $15\,\hmpc$.

After applying the same isolation criteria of \citet{kauffmann13} to the halo quenching mock, we demonstrate
that the halo quenching model correctly reproduces the level of large-scale conformity in SDSS, as measured by
the red galaxy fractions around the red vs.\ blue primary galaxies. Confirming the results from
\citet{tinker17} and~\citet{sin17}, we find that ${\sim}19\%$ of the primaries in the mock are mis-identified
satellite galaxies, which contribute a significant false conformity signal to the red fraction of
distant neighbours.

To summarize, the fiducial halo quenching model provides a remarkably simple yet accurate picture of the
spatial distribution of galaxy colors in the local Universe~($z\,{<}\,0.25$), including the colour dependence
of galaxy clustering and weak gravitational lensing, the mark correlation functions of colours, and the
large-scale galactic conformity in SDSS. On the contrary, models that rely on the halo assembly bias effect to
quench star formations have great difficulties in matching to those observations, whether it be the
age-matching model of \citeauthor{hearin14} or the assembly-biased halo quenching model developed in our analysis.

\subsection{Alternative Environmental Effects as a Test of Halo Quenching}

There are at least two aspects of the large-scale environment that one could imagine might affect halo and
galaxy properties:
one is the large-scale background density, and the
other is the large-scale tidal tensor field, i.e., the geometrical environment. In this paper we are primarily
concerned with the former, but there could be additional dependences of quenching on the geometrical
environment, which can be classified as clusters, filaments, sheets and voids via the Hessian matrix of the
gravitational potential~\citep{hahn07, forero-romero09, hoffman12}. \citet{alonso15} investigated the
variation of halo mass functions in different geometrical environments and found that at fixed large-scale
overdensity, the halo mass functions are similar among the four different types of structures within the
cosmic web. Therefore, under the fiducial picture of halo quenching there should be no colour dependence on
the geometrical type of the environment.

However, some halo properties, like the halo spin and shape, depend strongly on the tidal and velocity shear
fields~\citep{hahn07}, and it is plausible that they could also affect galaxy quenching.  Observationally,
there is tentative evidence for correlations between galaxy properties~(colour, size, spin, morphology, bar
strength, etc) and the proximity to various geometrical environments~\citep{pandey06, zhang13, zhang15,
chen17}. Therefore, it will be interesting to predict the distributions of galaxy colours in different
geometrical environments from the fiducial halo quenching mock, and compare to the measurements from data. Any
discrepancies would signal additional environmental effects that cannot be accounted for by the halo quenching
model, or any galaxy formation models that are insensitive to the tidal tensor field.

% Beside the sharp changes in their spectral energy distributions, galaxies also experience morphological and
% structural transformations during the quenching phase~\citep{kauffmann03, bell12, fang13, woo17}. It remains
% uncertain whether the structural change is one of the drivers of quenching~\citep{martig09}, or merely a
% consequence of the redshift evolution of galaxy sizes that are unrelated to quenching physics~\citep{lilly16}.

\subsection{Theoretical Implications and Future Outlook}

Under our fiducial halo quenching picture, the efficiency of galaxy quenching is tied to the strength
of the gravitational potential in the host halos, which could drive virial shocks that heat up the gas
and/or harbor AGNs that inhibit star formation via powerful feedback processes. The great success of
this simple picture, combined with the lack of any discernible galaxy assembly bias effect, suggests
that galaxy quenching is likely a local process, in the sense that it is largely determined by the
physical conditions inside a halo.  In particular, the termination of star formation becomes prevalent
when the halo reaches a critical mass of $M_h^{\mathrm{crit}}\,{\sim}\,10^{12}\,\hmsol$. This value
of $M_h^{\mathrm{crit}}$, derived from the observational constraints in Paper II and in this paper
under a different cosmology, is roughly the same for central and satellite galaxies, and is naturally
expected from canonical halo quenching theory~\citep{birnboim03, dekel06, cattaneo06}. This scenario of
galaxy quenching being locally driven by host halos is also consistent with the lack of environmental
dependence in the mass-metallicity relation~(MZR) of galaxies~\citep{wu17},
while a strong correlation between quenching and halo formation time would instead shift the
the overall amplitude of the MZR with the large-scale overdensity.

The ability of our fiducial halo quenching model to capture the observed environmental dependence
of colours is encouraging news to the modelling of large-scale structure in ongoing and upcoming
surveys.  In particular, the construction of large mock galaxy catalogues with realistic stellar mass
and colour properties is vital to the success of future surveys like the Dark Energy Spectroscopic
Instrument~\citep[DESI;][]{desi16}, Prime Focus Spectrograph~\citep[PFS;][]{takada14}, and Large
Synoptic Survey Telescope~\citep[LSST;][]{lsst09}, for validating analysis pipelines and extracting
cosmological information~\citep[see also][]{smith17}. For example, the \ihod\ mock galaxy catalogue is
one of the primary synthetic sky catalogues that are deployed within the web-based mock catalog validation and comparison framework
for the LSST Dark Energy Science Collaboration~(DESCQA; Heitmann et al, {\it in prep}). Furthermore, the success of the halo quenching
model reinforces the theoretical foundation for the red-sequence based cluster finding algorithms in
modern photometric surveys~\citep[][]{rykoff14} --- the richness of a cluster $\lambda$, defined by
$\lambda(\mh) \equiv f_{\mathrm{red}}(\mh) \times N_{\mathrm{sat}}(\mh)$,
% \begin{equation}
% \lambda(\mh) \equiv f_{\mathrm{red}}(\mh) \times N_{\mathrm{sat}}(\mh),
% \end{equation}
is more tightly correlated with $\mh$ than either $f_{\mathrm{red}}$ or $N_{\mathrm{sat}}$ individually,
and is insensitive to the average age of its subhalos or the large-scale environment.

Another exciting prospect is to apply the \ihod\ halo quenching framework to high redshifts.  In particular,
the Bright Galaxy Survey~(BGS) program within DESI will conduct a magnitude-limited survey of approximately
10 million galaxies with a median redshift of $0.2$. When combined with the current analysis of the SDSS
main spectroscopic sample, the \ihod\ modelling of the BGS galaxies will enable an exquisite understanding
of the evolution of galaxy quenching properties over the past ${\sim}2$~Gyrs, which is comparable to the
observed star formation efficiency timescale of the molecular gas~\citep{leroy08}, as well as the expected
timescale for quenching~\citep{wetzel13, hahn16}. At slightly higher redshifts~($z\,{\sim}\,0.5$),
a comprehensive \ihod\ modelling will also shed important insight on the apparent inconsistency
between the clustering and lensing of galaxies under the Planck15 cosmology~\citep{planck15, leauthaud16}.

Finally, this paper concludes our three-paper series by building the fiducial halo quenching mock catalogue
that accurately reproduces the spatial clustering, weak lensing, and density environment of galaxies at any
given stellar mass and colour within the local Universe.  We will be happy to provide our mock catalogues or
generate new mock catalogues with user-defined halo catalogues for those interested, upon request.

\section*{Acknowledgements}

We thank Ravi Sheth and David Weinberg for helpful discussions. YZ and RM acknowledge the support by
the U.S. Department of Energy~(DOE) Early Career Program. YZ is also supported by a CCAPP fellowship.
The authors gratefully acknowledge the Gauss Centre for Supercomputing e.V. (www.gauss-centre.eu)
and the Partnership for Advanced Supercomputing in Europe (PRACE, www.prace-ri.eu) for funding the
MultiDark simulation project by providing computing time on the GCS Supercomputer SuperMUC at Leibniz
Supercomputing Centre (LRZ, www.lrz.de).  The Bolshoi simulations have been performed within the Bolshoi
project of the University of California High-Performance AstroComputing Center (UC-HiPACC) and were run
at the NASA Ames Research Center.



% \clearpage
\bibliographystyle{mnras}
\bibliography{bibliography/biblio} %

% \begin{thebibliography}{99}
% \bibitem[\protect\citeauthoryear{Author}{2012}]{Author2012}
% Author A.~N., 2013, Journal of Improbable Astronomy, 1, 1
% \bibitem[\protect\citeauthoryear{Others}{2013}]{Others2013}
% Others S., 2012, Journal of Interesting Stuff, 17, 198
% \end{thebibliography}

%%%%%%%%%%%%%%%%%%%%%%%%%%%%%%%%%%%%%%%%%%%%%%%%%%

% Don't change these lines
\bsp	% typesetting comment
\label{lastpage}

\end{document}